\documentclass[useAMS]{mn2e}

\def\dg{$^{\circ}$}
\def\msun{\hbox{$M_{\odot}$}~}

\def\teff{${T}_{\rm eff}$}
\def\kms{{km~s}$^{-1}$}
\def\logg{$\log g$}
\def\turb{$\xi$}
\def\rad{$v_{\rm r}$}
\def\vsini{$v$ sin $i$}
\def\rsun{$R_{\odot}$~}

\usepackage{graphicx}
 \usepackage{times}
  \usepackage{rotating}
  \usepackage{txfonts}

\title{Fundamental parameters of the close interacting binary HD\,170582 and its luminous accretion disc}
\author[Mennickent et al.]
  {R.E. Mennickent$^{1}$\thanks{E-mail: rmennick@astroudec.cl },  
     G. Djura\v{s}evi\'c$^{2,3}$, 
   M. Cabezas$^{1}$,
  A. Cs\'eki$^{2}$,
   J. Rosales G.$^{1}$,
   \newauthor
    E. Niemczura$^{4}$, 
    I. Araya$^{5}$,
      M. Cur\'e$^{5}$
\\
  $^1$Universidad de Concepci\'on, Departamento de Astronom\'{\i}a,
      Casilla 160-C, Concepci\'on, Chile\\
  $^{2}$ Astronomical Observatory, Volgina 7, 11060 Belgrade 38, Serbia   \\
   $^{3}$ Isaac Newton Institute of Chile, Yugoslavia Branch\\
    $^4$  Astronomical Institute, Wroc{\l}aw University, Kopernika 11, 52-622
Wroc{\l}aw, Poland\\
  $^{5}$ Instituto de F\'{\i}sica y Astronom\'{\i}a, Facultad de Ciencias, Universidad de Valpara\'{\i}so, Chile 
  }
\date{}

\pagerange{\pageref{firstpage}--\pageref{lastpage}} \pubyear{2008}

\def\LaTeX{L\kern-.36em\raise.3ex\hbox{a}\kern-.15em
    T\kern-.1667em\lower.7ex\hbox{E}\kern-.125emX}

\begin{document}

\label{firstpage}

\maketitle 

\begin{abstract} 
We present a spectroscopic and photometric study of the Double Period Variable HD\,170582. 
Based on the study of the ASAS V-band light curve we determine 
an improved orbital period of  16.87177 $\pm$ 0.02084 days and a long 
period of 587 days.  We disentangled the light curve into an orbital part, determining ephemerides and revealing orbital ellipsoidal variability with unequal maxima, and a long cycle, showing quasi-sinusoidal changes with amplitude $\Delta V$= 0.1 mag. Assuming synchronous rotation for the cool stellar component and semi-detached configuration we find a cool evolved star of $M_{2}$ = 1.9 $\pm$ 0.1 $M_{\odot}$, $T_{2}$ = 8000 $\pm$ 100 $K$ and $R_{2}$ = 15.6  $\pm$ 0.2 $R_{\odot}$, and an early B-type dwarf of $M_{1}$ =  9.0 $\pm$ 0.2 $M_{\odot}$. 
The B-type star is surrounded by a geometrically and optically thick accretion disc of radial extension  
20.8 $\pm$ 0.3 \rsun  contributing about 35\% to the system luminosity at the $V$ band. Two extended regions located at opposite sides of the disc rim, and hotter than the disc by 67\% and 46\%, fit the light curve asymmetries.   
The system is seen under inclination 67.4 $\pm$ 0.4 degree and it is found at a distance of 238 $\pm$ 10 pc. 
Specially interesting is the double line nature of \textsc{He\,i} 5875; two absorption components move in anti-phase during the orbital cycle; they can be associated with the shock regions revealed by the photometry. The radial velocity of one of the \textsc{He\,i} 5875 components closely follows the donor radial velocity,
suggesting that the line is formed in a  wind emerging near the stream-disc interacting region. 

\end{abstract}

\begin{keywords}
stars: early-type, stars: evolution, stars: mass-loss, stars: emission-line,
stars: variables-others
\end{keywords}

\section{Introduction}

HD\,170582 (BD-14 5085, ASAS ID 183048-1447.5, $\alpha_{2000}$ = 18:30:47.5, $\delta_{2000}$ = -14:47:27.8, $V$ = 9.66 mag, $B-V$ = 0.41 mag, spectral type A9V)\footnote{http://simbad.u-strasbg.fr/simbad/}
is a poorly studied binary star catalogued ESD (semi-detached eclipsing binary) and with orbital period 16.8599 days in the ASAS\footnote{http://www.astrouw.edu.pl/asas/}  catalogue (Pojma\'nski 1997).  
It is located in the region of the cool molecular cloud L\,379 and was observed by Lahulla and Hilton (1992) who obtained  $V$ = 9.62 mag, $B-V$ = 0.44 mag and $U-B$ = -0.27 mag. 
The system is characterized by a long photometric cycle of 536 days  and is the third longest-period member of the Galactic Double Period Variables (DPVs), after V\,495 Cen and V\,4142 Sgr (Mennickent \& Rosales 2014, Mennickent et al. 2012a). DPVs are intermediate mass interacting binaries showing a long photometric cycle lasting about 33 times the orbital period, which has been interpreted as cyclic episodes of  mass loss (Mennickent et al. 2003, Mennickent et al. 2008, 2012b, Poleski et al. 2010).  More than 200 DPVs have been found in the Galaxy and the Magellanic Clouds (Mennickent 2013), but very few of them have been studied spectroscopically (e.g. Barr{\'{\i}}a et al. 2013, Garrido et al. 2013).  
The study of HD\,170852 is important to characterize DPVs in terms of their fundamental physical parameters and also to help to understand the still unknown cause for the long photometric cycle.

In this paper we determine fundamental orbital parameters and physical parameters for the stellar components and for the accretion disc surrounding the more massive star.  We use indistinctly the terms primary or gainer for the more massive star and
secondary or donor for the less massive star. The analysis of the circumstellar matter, long cycle and evolutionary stage are postponed for a forthcoming paper. In Section 2 we present the analysis of the ASAS light curve and derive photometric ephemerides. In Section 3 we present our high-resolution spectroscopy which is analyzed in Section 4 determining basic parameters for the cool stellar component and the system mass ratio. In Section 5 we model the light curve with a special code including light contributions of both stars and the accretion disc, determining stellar temperatures, radii, luminosities, surface gravities and masses and the system inclination. The characteristics of the accretion disc are  determined in Section 5. The spectral energy distribution is analyzed in Section 6, determining reddening and distance. We end in Section 7 summarizing the main results of our research. 

\section{photometric ephemerides}

We re-analyzed  the ASAS light curve considering only those better-quality data points labeled as A-type and B-type and  rejecting outliers and a cloud of deviating points around HJD\,2\,452\,471. The analysis was made on the remaining
455 data points.
The period searching algorithm PDM (Stellingwerf 1978) was used on the dataset, 
revealing an orbital period $P_{\rm{o}}$ = 16.87177 $\pm$ 0.02084 (the error corresponds to the half-width-at-half-minimum of the periodogram's peak)
and epoch of maximum HJD = \,2\,452\,118.2751 $\pm$ 0.337 days. 
A second periodicity was detected, $P_{\rm{l}}$  = 587 days, with a broad asymmetrical periodogram's peak, characterized by a full-width-at-half-minimum of 85 days and epoch for maximum  HJD\,2\,452\,070.88 $\pm$ 17.61 days. We noticed that the ASAS period (16.8599 days) does not fit the periodogram minimum as well as our period, probably because of the automatic character of the
period searching algorithm used in this catalogue and the lack of filtering of bad data points. The light curve was disentangled with these two periods using the software written by Zbigniew Ko{\l}aczkowski and described in Mennickent et al. (2012a). Afterwards, the resulting disentangled light curves  were folded with both periods as shown in Fig.\,1,  revealing an orbital modulation typical for an ellipsoidal binary but with unequal maxima and longer cycle characterized by a quasi-sinusoidal variability. The difference in maxima in the orbital light curve could indicate a non-axisymetrical brightness distribution in the orbital plane.

\begin{figure}
\scalebox{1}[1]{\includegraphics[angle=0,width=8.5cm]{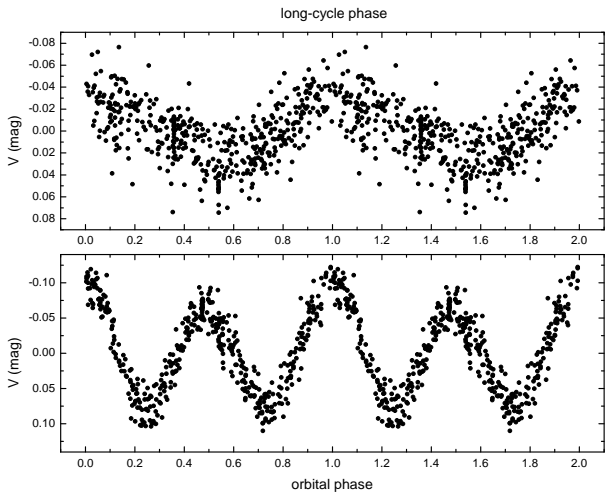}}
\caption{The disentangled ASAS light curves of HD\,170582 folded with the long period (up) and
the orbital period (down). Phases are calculated according to times of light curve maxima, given by Eqs.\,1 and 2.}
  \label{x}
\end{figure}

By phasing the disentangled light curves with their respective periods, and measuring the phases of maxima with polynomial fits,  
we determined the following ephemerides for the maxima of the light curves:\\

$HJD_{\rm{max, long}}  = 2\,452\,070.9 + 587 \times E, $\hfill(1)

$HJD_{\rm{max, orbital}}  = 2\,452\,118.275(34) + 16.871(21) \times E. $ \hfill(2) \\

These are used for the spectroscopic analysis and discussed
with the new spectroscopic ephemerides determined in Section 4.

\section{Spectroscopic Observations}

We conducted spectroscopic observations of HD\,170582 since
year 2008 to 2013 obtaining 13 optical  spectra with resolution $R \sim$ 40,000 with  the spectrograph CORALIE (La Silla ESO Observatory), 
112 spectra with CHIRON spectrograph with $R \sim$ 30,000  (fiber mode, Cerro Tololo Inter-American Observatory, CTIO) and 11 spectra with the DuPont-echelle spectrograph with  $R \sim$ 40,000 (Las Campanas Observatory, LCO)\footnote{Technical descriptions for these spectrographs and their cameras can be found in www.eso.org/ , http://www.ctio.noao.edu/ and http://www.lco.cl/}.
The spectral regions covered were 3865-6900 \AA~ (CORALIE), 4580-8760 \AA~ (CHIRON) and 3600-9850 \AA~  (DuPont-echelle). 



All spectra discussed in this paper are  normalized to the continuum and 
the RVs are heliocentric ones. Reductions were done with IRAF\footnote{IRAF is distributed by the National Optical Astronomy Observatories,
 which are operated by the Association of Universities for Research
 in Astronomy, Inc., under cooperative agreement with the National
 Science Foundation.} following usual procedures for echelle spectrography, including flat and bias correction, wavelength calibration and order merging. 
As a measure of internal error of the wavelength calibration we measured the position of the interstellar Na\,D1 line with $rms$ accuracy of 0.5 km s$^{-1}$.   


The spectra obtained with the optical fiber spectrographs CORALIE and CHIRON are not sky-subtracted.  This limitation  has no effect for radial velocity  and  line strength measurements, since HD\,170582 is  bright even at full moon and we do not flux-calibrate our spectra. Details for our observational runs  are given in Table\,1. 

\section{Spectroscopic analysis}

\subsection{Determination of donor physical parameters}

 The first inspection of the spectroscopic material reveals a SB2 type binary consisting of a late-A star with sharp metallic lines and a less luminous B star  with broader helium absorption lines. In addition, emission in Balmer lines is observed at some epochs suggesting  that the system is an  interacting binary.  By comparison of relative strengths of metallic lines in the region 4500-4600 \AA, devoid of lines of the hotter component, with some spectra from the UVES-POP library\footnote{https://www.eso.org/sci/observing/tools/uvespop.html}, we find a relatively good fit with the diluted spectrum of HD\,90772, hence we estimate a spectral type A9 for the cooler component, in agreement with the figure given by Houk and Smith-Moore (1988). Comparing luminosity-sensitive features like the \textsc{Fe\,ii} \& \textsc{Ti\,ii} double blend at  4172-8 \AA, and similar blends at 4395-4400 \AA, 4417 \AA~ and 4444 \AA, with less  sensitive luminosity lines like \textsc{Ca\,i} 4227, \textsc{Fe\,i} 4271 and \textsc{Mg\,ii} 4481, we estimate a luminosity class between I and III for the cooler star.

In order to determine the physical parameters of the A-type star, we compared our observed donor spectrum  with synthetic spectra constructed with the \textsc{synthe} code which uses atmospheric models computed with the line-blanketed LTE \textsc{atlas9} code (Kurucz 1993). The Kurucz's models are constructed with the assumptions of plane-parallel geometry and hydrostatic and radiative equilibrium of the gas. \textsc{Atlas9} was ported under GNU Linux by Sbordone \cite{sbordone} and is available online\footnote{http://atmos.obspm.fr/}. The use of a LTE grid for studying a A-type supergiant atmosphere, which could be affected by NLTE effects, could introduce an underestimation of the iron group abundances  by a factor of 2 to 3 (Przybilla et al. 2006).

The stellar line identification and the abundance analysis in the entire observed spectral range were performed on the basis of the line list from Castelli \& Hubrig (2004).\footnote{http://wwwuser.oat.ts.astro.it/castelli/grids.html}.
The theoretical models were calculated for effective temperatures from 7000 to 9000 $K$ with steps of 100 $K$, surface gravities from 1.0 to 3.5 dex with the step of 0.5 dex, solar metallicity and microturbulences from 0.5 to 3.0 with the step of 0.1 km/s. As a template we choose the average donor spectrum obtained after shifting all spectra to the donor system of rest, hence removing at first order the gainer contribution. The velocities used are derived in the next section.

Our analysis follows the methodology presented in Niemczura \&Po{\l}ubek (2006) and relies on an efficient spectral synthesis based on a least squares optimization algorithm. This method allows for the simultaneous determination of various parameters involved with stellar spectra and consists of the minimization of the deviation between the theoretical flux distribution and the observed normalized one. The synthetic spectrum depends on the stellar parameters, such as effective temperature \teff, surface gravity \logg, microturbulence \turb, rotational velocity \vsini, radial velocity \rad, and the relative abundances of the elements $\varepsilon_{El}$, where $El$ denotes the individual element. The first three parameters were determined before the determination of abundances of chemical elements. The \vsini\, value was determined by comparing the shapes of metal line profiles with the computed profiles, as shown in Gray\,(2005).

The effective temperature, surface gravity and microturbulence were determined by the analysis of neutral and ionized Fe lines. In this method we adjust \teff, \logg\ and \turb\ by the comparison of the abundances determined from \textsc{Fe\,i} and \textsc{Fe\,ii} lines. The analysis is based on iron lines because they are the most numerous in the spectrum. In general, we require that the abundances measured from \textsc{Fe\,i} and \textsc{Fe\,ii} lines yield the same result. The strength of absorption lines of \textsc{Fe\,i} depend on \teff, \turb\ and overall metallicity $Z$, and are practically independent from \logg. On the other hand the lines of \textsc{Fe\,ii} are slightly sensitive to the temperature, metallicity and most of all to gravity. First, we adjust the microturbulence until we see no correlation between iron abundances and line intensity for the \textsc{Fe\,i} lines. Second, \teff\ is changed until we see no trend in the abundance versus excitation potential of the atomic level causing the \textsc{Fe\,i} lines. Then, the gravity is obtained by fitting the \textsc{Fe\,ii} and \textsc{Fe\,i} lines and by requiring the same abundances from both neutral and ionized lines.

The \textsc{Fe\,i} abundances were calculated for a series of lines characterized by different excitation potential and line depth, in a grid of spectra spanning a range of values of log\,g, $T_{eff}$ and  \turb. From  a careful examination of these calculations  we estimated the mentioned parameters and their  uncertainties.

From the above analysis we obtained the best model characterized by  $T_{2}$ =  8000 $\pm$ 100 K, log$\,g_{2}$ = 1.7 $\pm$ 0.5 , $v_{\rm{2r}}$ sin $i$ = 44 $\pm$ 2 km s$^{-1}$ and \turb ~= 1.0 $\pm$ 0.7 km s$^{-1}$. An example of the fit of this model with the observed spectrum is shown in Fig.\,2.  

 For these parameters we obtained the abundances of chemical elements. We adopt the usual astronomical scale for logarithmic abundances where hydrogen is defined to be log $\epsilon_{H}$ = 12.00, i.e. log $\epsilon_{El}$ = log $(N_{El}/N_{H})$ + 12, where $N_{El}$ and $N_{H}$ are the number densities of element $El$ and hydrogen, respectively (Table\,2). They are average abundances, which means that for example for iron it is the average abundance obtained from \textsc{Fe\,i} and \textsc{Fe\,ii} lines. The \textsc{Fe\,i} abundance is equal $7.47\pm0.16$ (Fig.\,3). In this figure we show an histogram for the abundance obtained with different lines, the abundance versus line excitation potential and the abundance versus line strength, measured as the line depth. We used line depth instead equivalent width in order to include in the analysis some weakly blended lines. The use of the line depth is justified since in general, line depth and EW should be correlated to first order and since as the resolving power through the whole spectral range is practically constant (R $\approx$ 30\,000), we should expect, for a given EW, and increase of line width of 30\% between 4000 and 5000 \AA. This means a decrease of line depth by the same order of magnitude, that should be translated into small horizontal shifts of the points in the right graph of Fig.\,3, without affecting the general pattern. This will not happen  if the line is saturated. If the line is on the flat or damping portion of the curve-of-growth the assumption will fail. According to the obtained \textsc{Fe} abundance the metallicity of the star is approximately solar (Table\,2 and Fig.\,4). 

 In some Algols, the donor has transferred important part of its atmosphere into the gainer, exposing its inner layers rich in elements produced by thermonuclear fusion during the star's main-sequence evolution. In these cases  abundance analysis of the donor surface layers should reveal an excess of nitrogen and a carbon depletion  (e.g. Kolbas et al. 2014 and reference therein). Our analysis shows carbon with solar abundance but with a large error, based only on 3 lines.
 On the other hand, we find oxygen and barium under-abundant and sodium and cobalt overabundant. While the case of Ba might be affected by the absence of hyperfine structure in the analysis; the interpretation of the other discrepancies  is at present unclear.

\begin{figure*}
\scalebox{1}[1]{\includegraphics[angle=0,width=18cm]{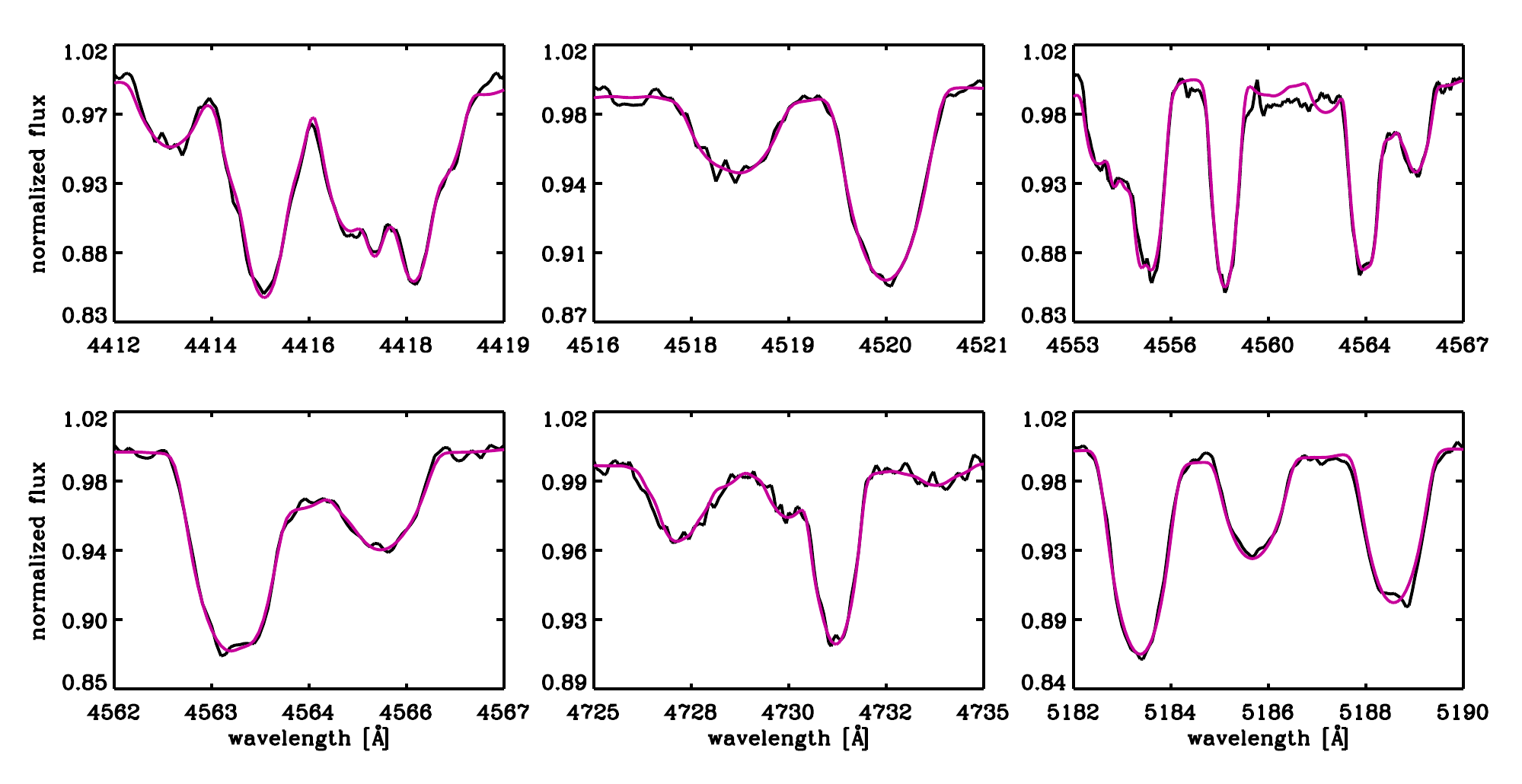}}
\caption{ Panels showing the detailed comparison between the observed and synthetic (best model; smoothed line) donor spectrum at different spectral ranges.}
 \label{x}
\end{figure*}

\begin{figure*}
\scalebox{1}[1]{\includegraphics[angle=90,width=18cm]{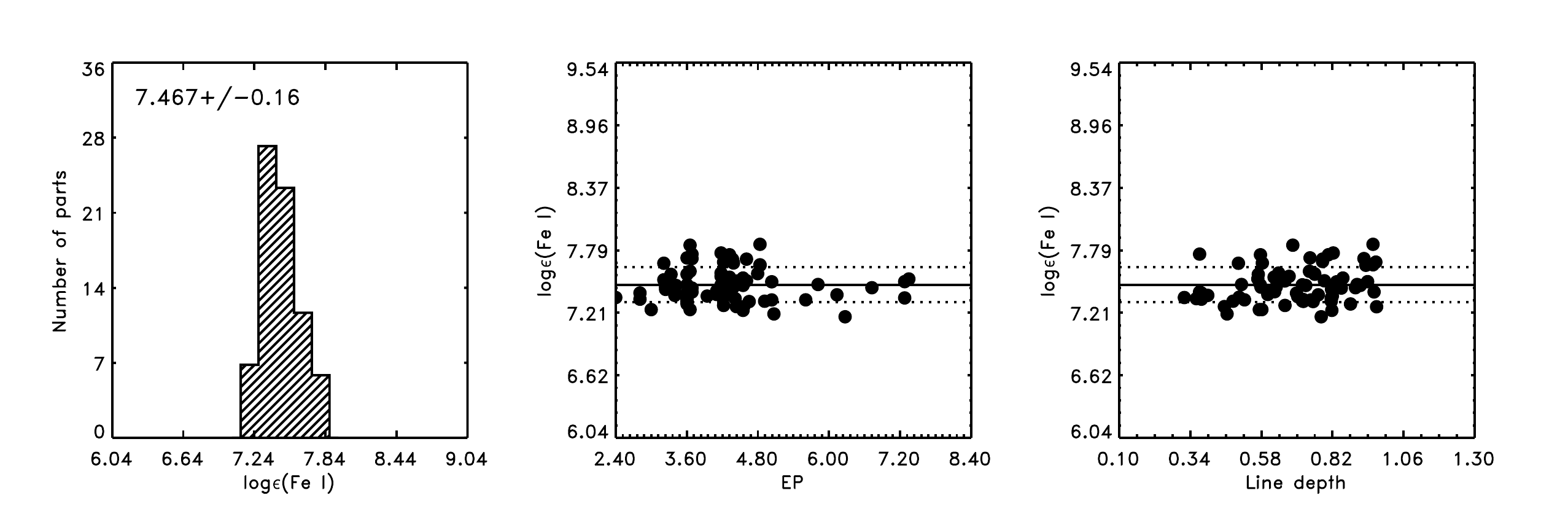}}
\caption{ Example of our calculations of \textsc{Fe\,i} abundance for log\,g= 1.7, \turb ~= 1.0 and $T_{eff}$ = 8000 K for lines of different excitation potential (EP) and depth. Left panel shows the average value and standard deviation. }
 \label{x}
\end{figure*}

\begin{figure*}
\scalebox{1}[1]{\includegraphics[angle=0,width=18cm]{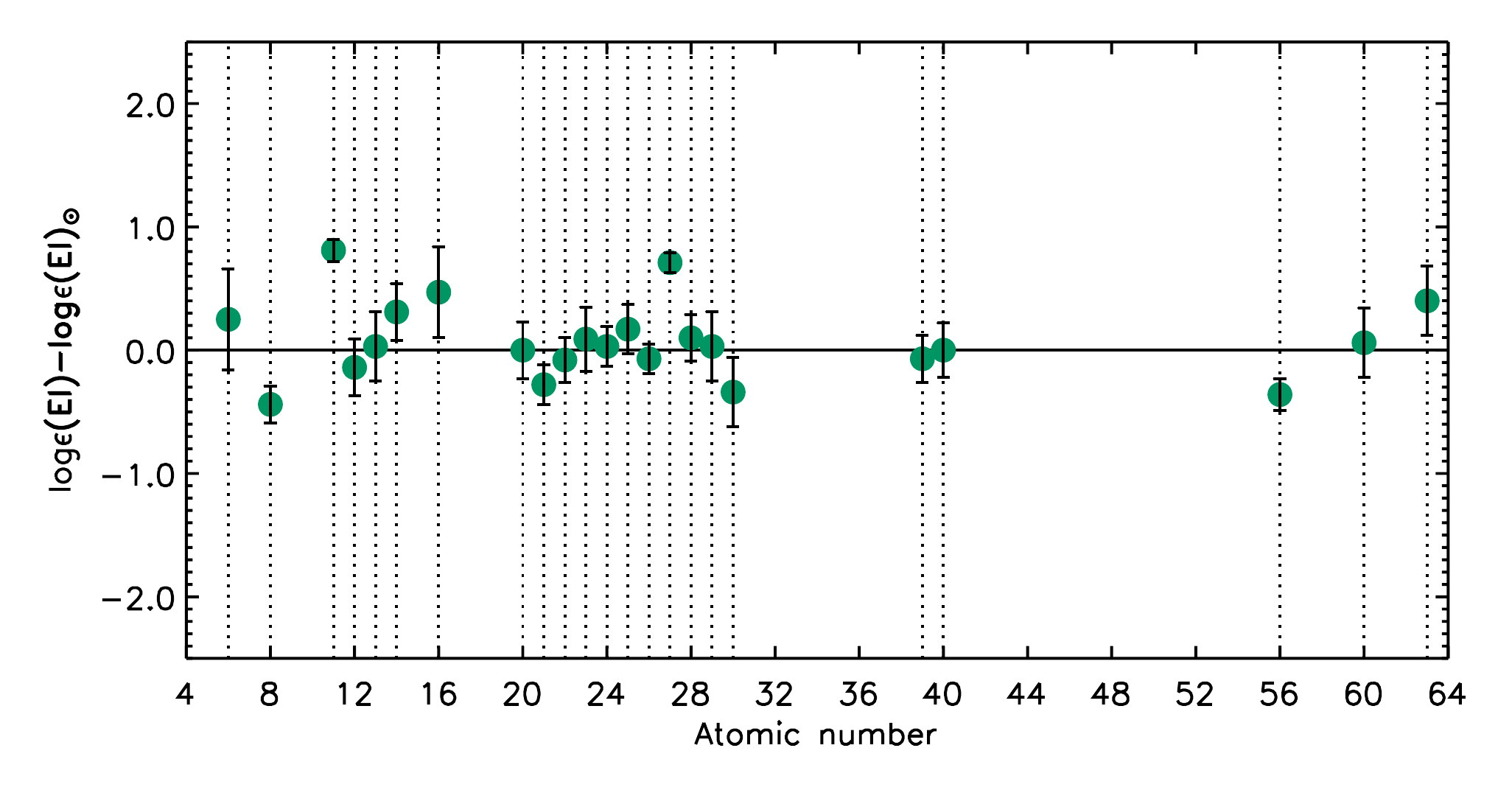}}
\caption{ Comparison between the abundances obtained from different elements and the solar values from Asplund et al. (2009).}
 \label{x}
\end{figure*}

\subsection{Radial velocities for the donor}

Radial velocities (RVs) for the donor were measured by cross-correlating the observed spectra with a reference spectrum and then applying the velocity shift corresponding to the template velocity. This last was obtained by fitting simple gaussians to some metallic lines, finding the central wavelength and comparing these wavelengths with the corresponding laboratory wavelengths. The cross-correlation was performed in  two regions deployed of \textsc{H\,i} and \textsc{He\,i} lines, viz.\,4500--4800 \AA~ and 5050--5680 \AA.  
The radial velocities are given in Table\,3.  Subsequent inspection revealed 
much larger scatter in the RVs of spectra taken at LCO and ESO, they were not considered in the following analysis, i.e. we give more confidence to CHIRON based velocities and they are used in the rest of the paper. 

The RVs can be fitted with a sinusoid with radial-velocity  half-amplitude $K_{2}$ = 139.8 $\pm$ 0.2 km s$^{-1}$ and  zero point  $\gamma$ = -1.3 $\pm$ 0.2 km s$^{-1}$.  A careful inspection of residuals shows a non-random distribution for the circular fit.

 In order to resolve the question about the possible ellipticity of the orbit, we used the genetic algorithm \textsc{pikaia} developed by Charbonneau (1995) to find the orbital elements for
HD\,170582. The method consists in finding the set of orbital parameters that produces a series of theoretical velocities that minimize the  function $\chi^{2}$ defined as:\\

\begin{figure}
\scalebox{1}[1]{\includegraphics[angle=0,width=8.5cm]{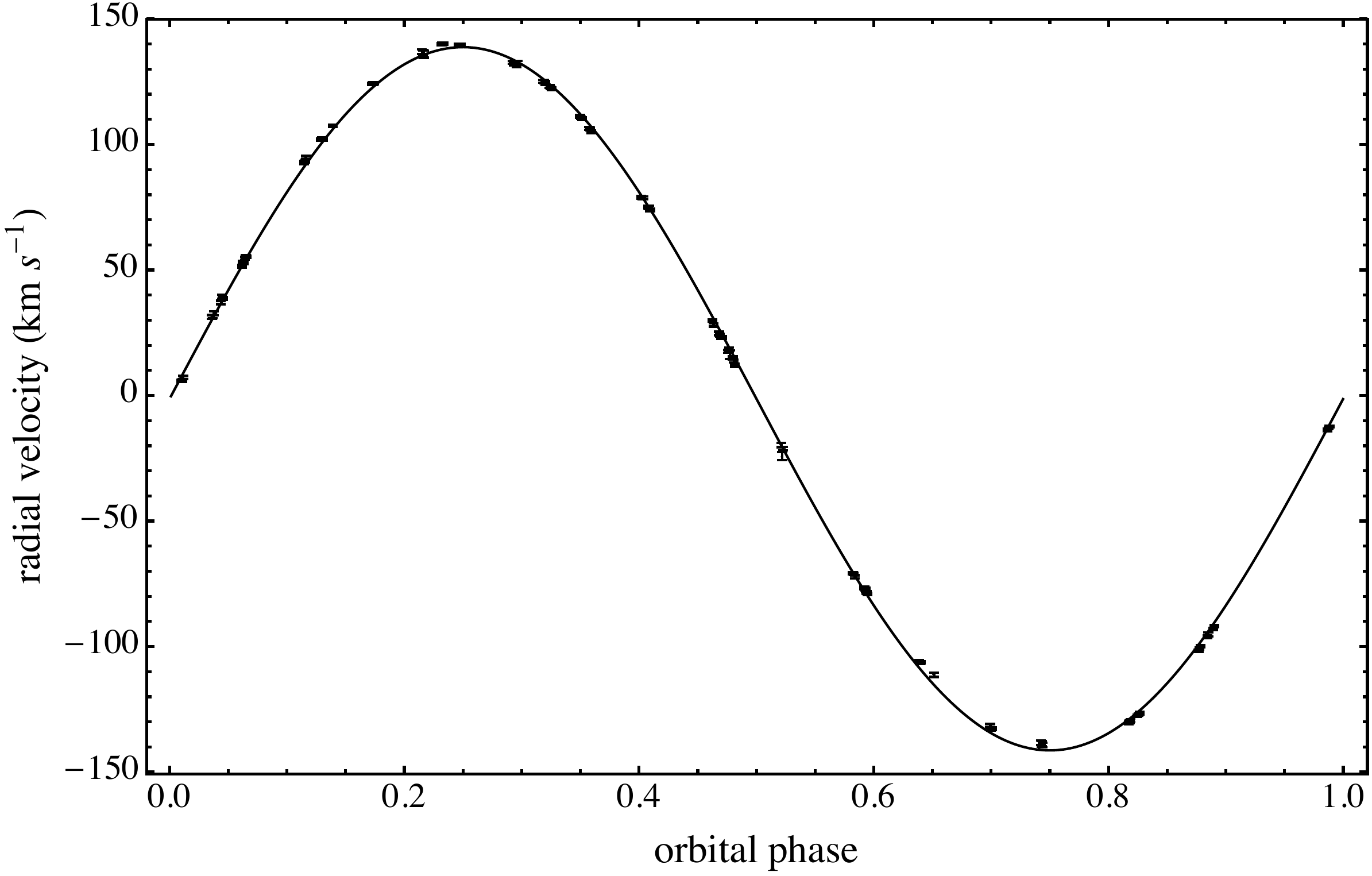}}
\scalebox{1}[1]{\includegraphics[angle=0,width=8.5cm]{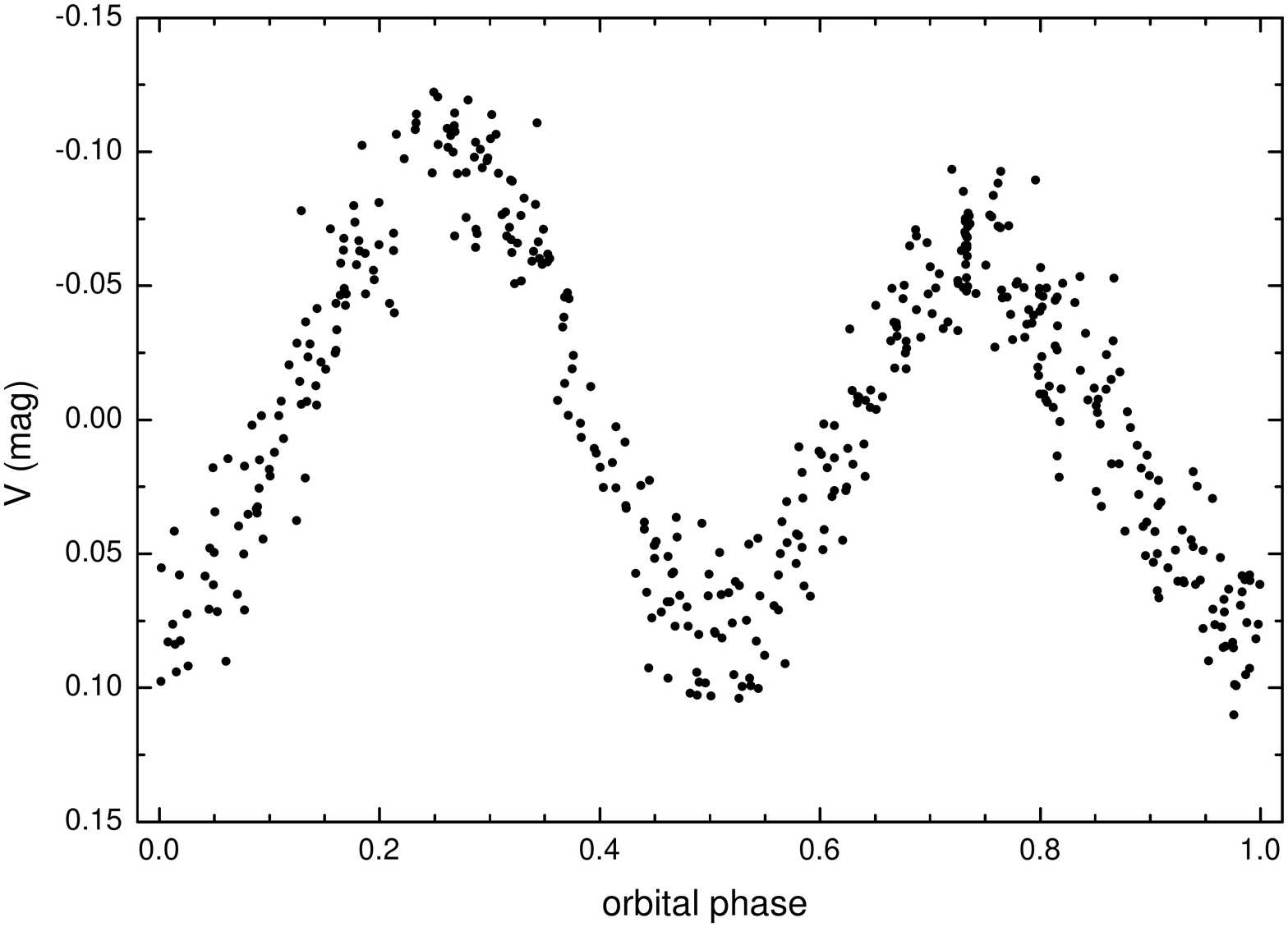}}
\caption{Upper panel: the donor radial velocities and the best fit, given by Eq.\,4 and the parameters of Table\,4. The radial velocity error bars are also shown, but they are usually smaller than the used symbols. Lower panel: disentangled orbital light curve. In both panels phases are calculated according to times of  donor inferior conjunction, given by Eq.\,7. }
  \label{x}
\end{figure}

\begin{small}
$\chi^{2} (P_{\rm{o}}, \tau, \omega, e, K_{2}, \gamma) = $\\

$\frac{1}{N-6} \displaystyle\sum\limits_{j=1}^N 
(\frac{V_{j}-V(t_{j}, P_{\rm{o}}, \tau, \omega, e, K_{2}, \gamma)}{\sigma_{j}}), 
$\hfill(3)\\
\end{small}

\noindent
where $N$ is the number of observations, $P_{\rm{o}}$ is the orbital period, $\omega$ the periastron longitude, $\tau$ the time of passage per the periastron, $e$ the orbital eccentricity, $K_{2}$ the half-amplitude of the radial velocity for the donor and $\gamma$ the velocity of the system center of mass. $V_{j}$ and $V$ are the observed and theoretical radial velocities at $t_{j}$. The theoretical velocity is given by:\\

$V(t) = \gamma + K_{2} (cos (\omega+\theta(t)) + e~ cos (\omega)), $\hfill(4)\\

\noindent
where $\theta$ is the true anomaly obtained  solving the following two equations involving the eccentric anomaly $E$:\\

$tan (\frac{\theta}{2}) = \sqrt{\frac{1+e}{1-e}}~ tan (\frac{E}{2}), $\hfill(5)\\

$E - e~sin(E) = \frac{2 \pi}{P_{\rm{o}}} (t - \tau). $\hfill(6)\\

A range of physically reasonable parameters need to be considered so that the method converge. For the period we used the range 10-20 days, the eccentricity was set between 0 and 1, $\omega$ between 0 and $2\pi$, $\tau$ between the minimum julian day and this value plus $P_{\rm{o}}$, $K_{2}$ between 0 and ($V_{\rm{max}} - V_{\rm{min}}$) and $\gamma$ between $V_{\rm{min}}$ and $V_{\rm{max}}$. 

The most reliable way to get error estimates for this genetic algorithm is by  Monte Carlo simulations, specifically by perturbing the best fit solution and computing the
$\chi^{2}$ of these perturbed solutions. To find the standard deviation region ($\sigma$)
 encompassed by the joint variation of two parameters with all other
parameters at their optimized values, we draw the contour corresponding to that value
of $\Delta \chi^2$ for 2 degrees of freedom that includes 68.3\% of the
probability. In our case this corresponds to $\Delta \chi^2 = 2.30$ 
(Bevington \& Robinson 1992, Chapter 11, p. 212).

The best orbital elements along with their estimated errors are given in Table\,4. The radial velocities and the best fit are shown  in Fig.\,5, along with the re-phased disentangled orbital light curve. We notice that the radial velocity and the light curve 
match the expectation for a close binary seen under an intermediate inclination, where the distorted stellar atmospheres show maximum projected surface (and brightness) just at times of radial velocity extremes. 
We find that the elliptical solution provides a much better fit than the circular case, since it gives residuals without systematic trends and also a smaller  $\chi^{2}$ value, viz.\, 2.39 versus 8.38. We note that our eccentric solution gives a small $e$ value (0.01) but it is highly significant, according to the statistical test ``$p_{1}$'' of Lucy (2005). In fact, following Lucy's definition, we calculated $p_{1}$= 2.5 $\times$ 10$^{-29}$ satisfying the condition less than 0.05 for a significant ellipticity.

Using the results above we calculated the time for the inferior conjunction of the donor 
finding the following ephemerides:\\

 $HJD_{0}  = (2\,456\,028.226 \pm 0.014)  + (16.8722 \pm 0.0017) \times E. $\hfill(7)\\
 
 \noindent
This ephemerides is used for orbital phases  in the rest of the paper whereas the phases for the long cycle refers to the ephemerides given by Eq.\,1.
 

 It has been pointed out that gas stream and circumstellar matter can distort spectroscopic features in semi-detached interacting binaries, producing skewed radial velocities and artificial small eccentricities (e.g. Lucy 2005). For a non-interacting binary with the stellar and orbital parameters of HD\,170582, dynamical tides should have circularized the orbit and synchronized the rotational  periods (Zahn 1975, 1977). This should imply that the observed small eccentricity can be spurious.
However,  as we will show later, the system is found with a circumprimary disc produced by mass transferred from the donor. It is  possible, but not here demonstrated, that the observed eccentricity could be the result of dynamical perturbations introduced by the accretion disc. If critical velocity is rapidly reached, as suggested by Packet (1981) and de Mink et al. (2007), then the disc could turn to be relatively massive, due to the inadequacy of the gainer of accreting more material. This last point has been debated; Petrovic et al. (2005) assume that accretion ceases when the mass gaining star reaches Keplerian rotation. On the contrary, Popham \& Narayan (1991) argue that a star near critical rotation can sustain accretion due to viscous coupling between the star and the disc.
In the case of HD\,170582, considering the luminous bright spots  reported in Section  5, the disc could be relatively massive with  asymmetrical mass distribution  and hence to produce a
small but non-zero orbital eccentricity.

\subsection{Gainer, mass ratio and circumstellar matter}

A gainer of B-type is suggested by the detection of \textsc{H\,i} and \textsc{He\,i}  absorption lines.  \textsc{Mg\,i\,4481} is dominated by the donor 
while the contribution of the gainer to this line is weak, if present; this indicates an early B-type for the gainer.  The helium lines
are contaminated by emission and show variable line profile shape, especially during high state. They are rather broad therefore the gainer might be
a rapidly rotating dwarf. The H$\alpha$ line is sometimes a sharp absorption with two cores, and the difference with the donor spectrum reveals a prominent double emission with deep central absorption (Fig.\,6).  This finding  supports the interacting binary nature for this system and suggests it is in a semi-detached stage.
\textsc{He\,i\,5875}  is very interesting since it shows two components moving in opposite directions (Fig.\,7).
During low stage ($\Phi_{\rm{l}}$ between 0.2 and 0.8) the radial velocity of the main \textsc{He\,i\,5875}  component C1 (after deblending by its nearby component)  as well as the \textsc{He\,i\,7065} line, can be fit with a  sine function of amplitude 75.1 $\pm$ 2.7 km s$^{-1}$ and zero point 20.6 $\pm$  2.3 km s$^{-1}$ (Fig.\,8). 
The secondary component C2 can be fit with a  sine of amplitud 127.2 $\pm$ 1.8 km s$^{-1}$ and zero point 11.8 $\pm$  1.4 km s$^{-1}$. This curve lags the donor RV curve only by  
$\Delta \Phi$ = 0.004 $\pm$ 0.003, i.e. it practically follows the donor motion. However, around $\Phi_{\rm{o}}$ = 0.75, the velocities turn to be less negative, which does not occurs around the other quadrature at $\Phi_{\rm{o}}$ = 0.25 (Fig.\,8). These velocities are given in Table\,5 and parameters for the RV fits are given in Table\,6.

If we use the system inclination $i = 67^{\circ}$ derived in Section 5, and the basic kinematics formula:\\

$\frac{r}{R_{\odot}} = \frac{Pv}{50.633} $ \hfill(8)\\

\noindent
where $r$ represents the radial distance from the center of rotation, $P$ the rotational period in days and $v$ the linear orbital velocity of material moving in the orbital plane measured in \kms, then we
can find the position of the light-centers of the line components listed in Table\,6. We find for C1 a light center located $r$ = 27.1  $R_{\odot}$ from the center of mass pointing 216$^{\circ}$ from the line joining the stellar centers measured   in the direction opposite to the orbital motion. For C2 we find the light center located at $r$ = 45.9 $R_{\odot}$ from the center of mass, pointing 1.4$^{\circ}$ from the line joining the stellar centers, as measured  in the direction opposite to the orbital motion. The meaning of these positions will be discussed  in Section 5.2.


From the above and also considering the irregular line profiles, it seems that the helium lines are not fully formed in the gainer stellar photosphere, but they can be partly formed in the circumstellar material. Actually, the overall helium lines cannot be fit by conventional synthetic line profiles of photospheric models; the line width changes and the depth is usually larger than expected for an early B-type gainer contributing about 50\% to the total light. This number comes from the veiling factor applied to the donor synthetic spectrum to match the observed metallic lines in Section 4.1. 
We considered the possibility that both components are artifacts produced by the motion of a central emission  feature, however in this case we should observe both absorption components
moving in phase which is not observed. 

Let's assume  for now  that the helium velocities of component C1 represent the gainer orbital motion, then the inferred mass ratio is $q$ = 0.54 $\pm$ 0.01.
We can investigate if this value is compatible with synchronous rotation for the secondary star. For a  secondary star filling its Roche-lobe in corotation with the binary:\\

$\frac{v_{\rm{2r}}sin\,i}{K_{2}} \approx (1+q)\frac{0.49q^{2/3}}{0.6q^{2/3}+\ln(1+q^{1/3})}$\hfill(9)\\
 
 \noindent
(Eggleton 2006, Eq. 3.9). 
Using the above equation, $K_{2}$ =  140.1 \kms and $v_{\rm{2r}}$ sin $i$ = 44 km s$^{-1}$, we obtain $q \approx$ 0.21, much lower than the $q$ value derived from helium lines (Fig.\,9). This indicates that the donor is rotating sub-synchronously or that the helium lines do not represent the motion of the gainer. We argue now against the first assumption. Synchronization time scales for early type stars are 100-1000 times shorter than circularization time scales (Hilditch 2001). As the system is almost circularized ($e$ = 0.013, Table\,4), there is no reason to suspect a non-synchronous donor. We are left with the explanation that C1 
is not formed in the gainer stellar photosphere. Their origin could be an accretion disc around the gainer, with asymmetrical brightness distribution, to account for the non-equal maxima in the light curve. 
We find support for this hypothesis in our light curve analysis presented  in Section 5.  
The mass ratio $q$  = 0.21 is favored in this paper, since it is consistent with donor spin synchronization and is also justified by arguments about  stellar masses and disc formation given in the next section.

\begin{figure}
\scalebox{1}[1]{\includegraphics[angle=0,width=9cm]{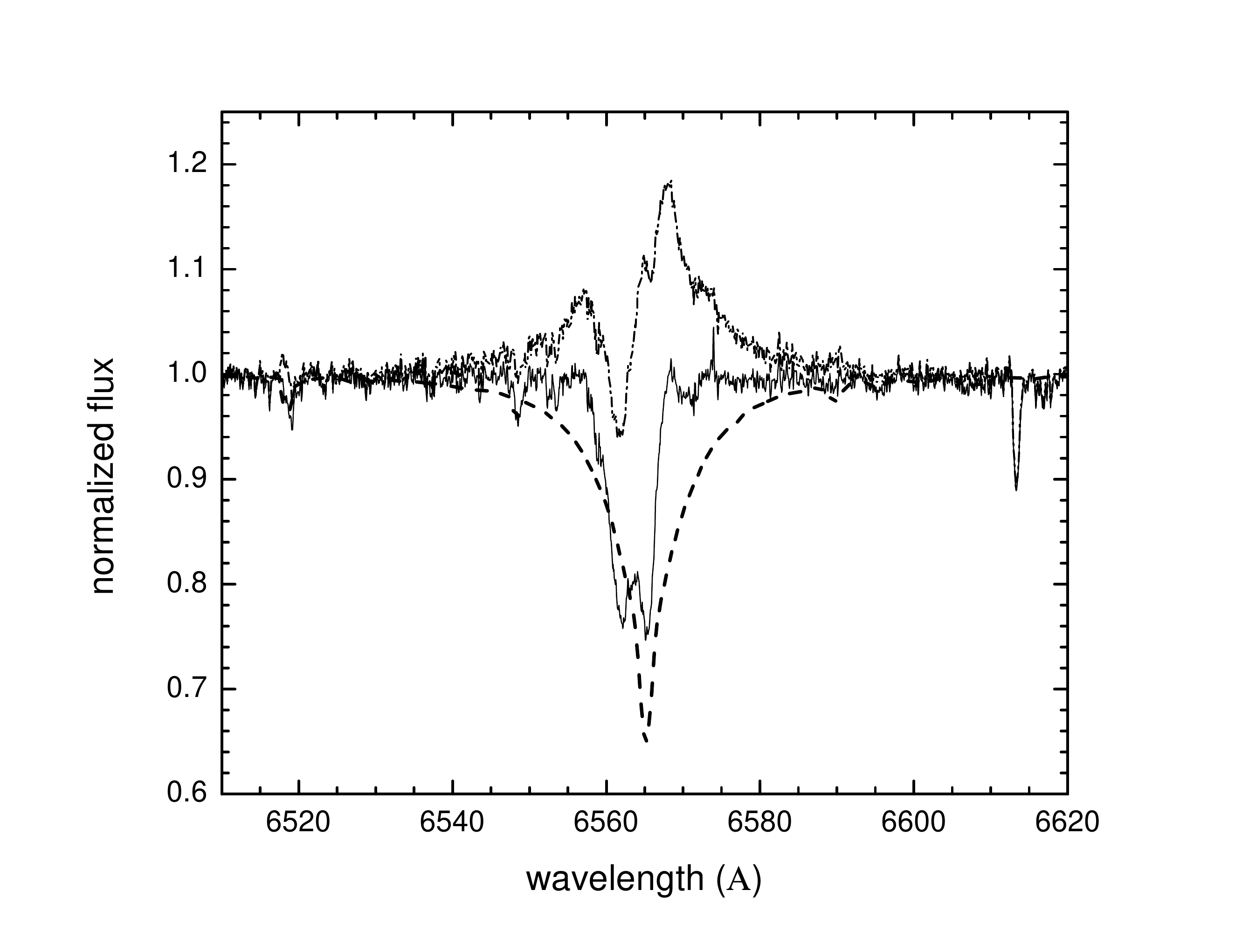}}
\caption{H$\alpha$ line on HJD\,2\,456\,506.680476 ($\Phi_{\rm{o}}$= 0.36, $\Phi_{\rm{l}}$= 0.56) over-plotted with the donor synthetic spectrum (thick dashed line) and the difference spectrum vertically shifted by +1.}
  \label{x}
\end{figure}

\begin{figure}
\scalebox{1}[1]{\includegraphics[angle=0,width=8.5cm]{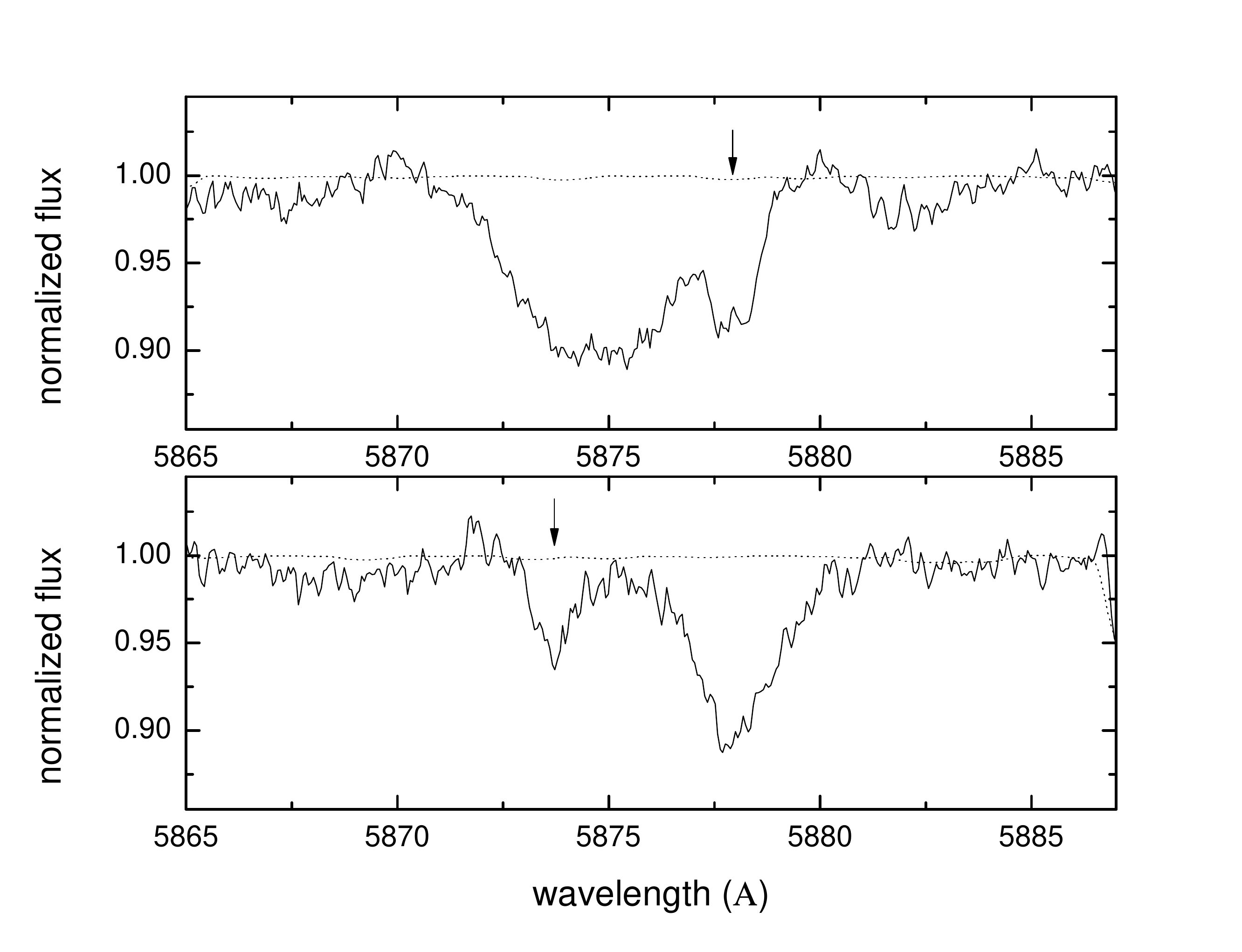}}
\caption{\textsc{He\,i\,5875}  lines for HJD\,2\,456\,489.690549 ($\Phi_{\rm{o}}$= 0.35, $\Phi_{\rm{l}}$= 0.53, up) and HJD\,2\,456\,497.705041 ($\Phi_{\rm{o}}$= 0.83, $\Phi_{\rm{l}}$= 0.54, down). The synthetic donor at the right velocity system is indicated with dotted lines and the arrow indicates the second \textsc{He\,i\,5875} component.}
  \label{x}
\end{figure}

\begin{figure}
\scalebox{1}[1]{\includegraphics[angle=0,width=9cm]{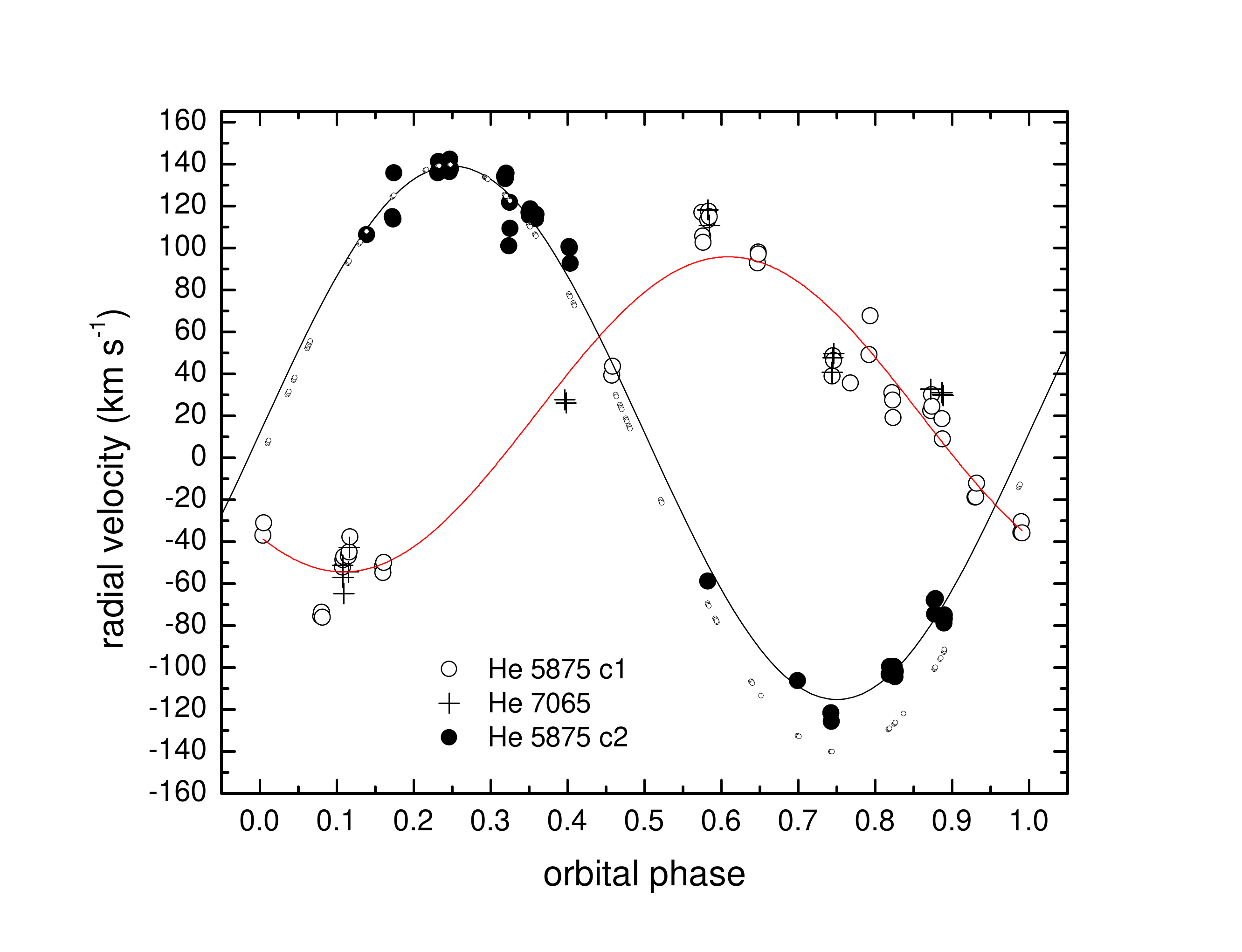}}
\caption{Radial velocity for the He\,5875 (components c1 and c2) and He\,7065 lines and the best  sine fits. Small circles show the donor RV at the observation epochs.   }
  \label{x}
\end{figure}

\begin{figure}
\scalebox{1}[1]{\includegraphics[angle=0,width=9cm]{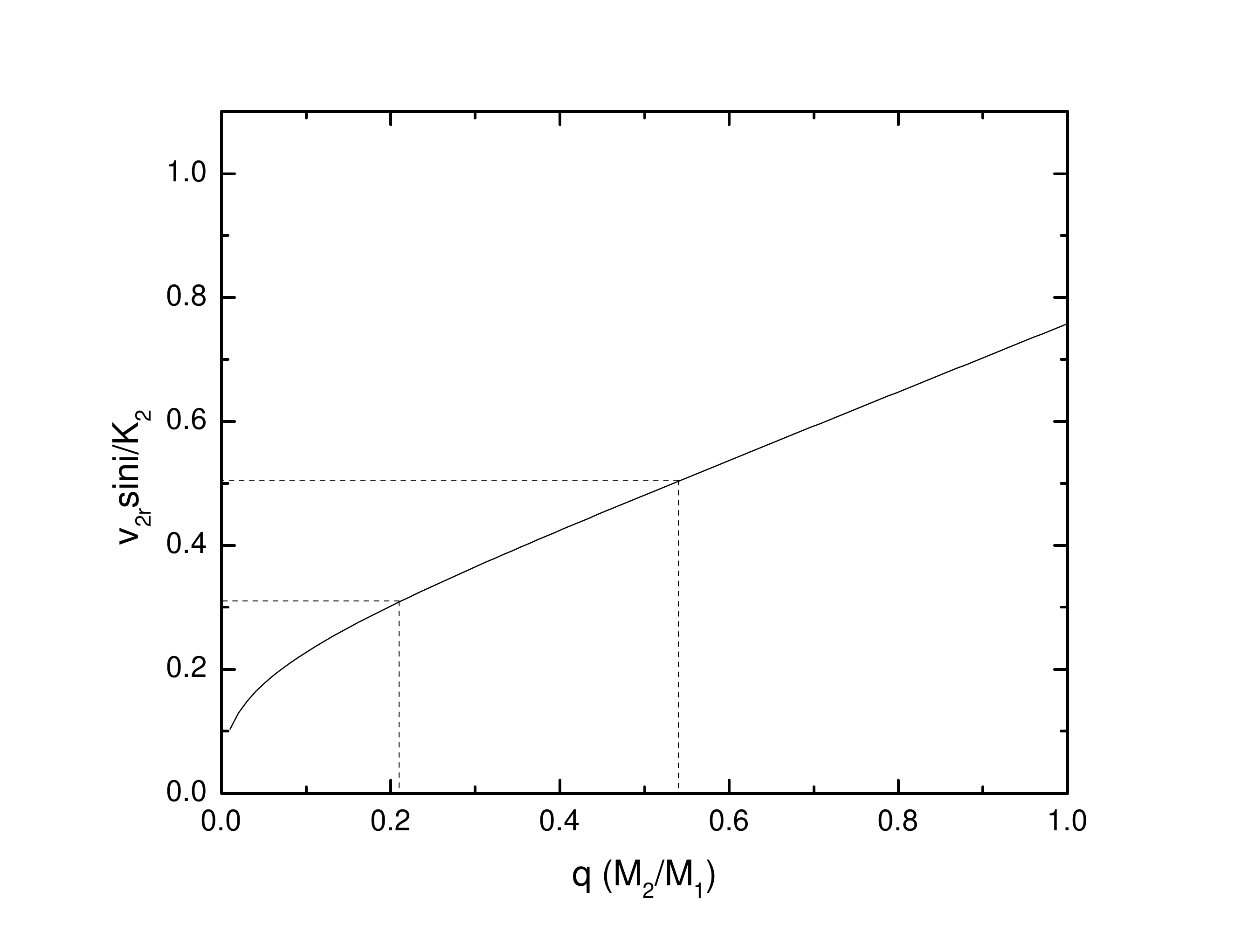}}
\caption{Relative donor rotational velocity versus mass ratio ($q$ = $M_{2}/M_{1}$). The solid line is given by Eq.\,(9) and the dashed lines show the synchronous ($q$ = 0.21) and the observed (sub-synchronous, $q$ = 0.54) cases.}
  \label{x}
\end{figure}



\subsection{Mass constrains from spectroscopy}

The system mass function for a binary in a circular orbit can be expressed as:\\

$f  = \frac{M_{2}sin^{3}i}{q(1+q)^{2}} = 1.0361\times10^{-7} (\frac{K_{2}}{km s^{-1}})^{3} \frac{P_{\rm{o}}}{day}$ M$_{\sun}$. \hfill(10) \\

\noindent 
The $f$ value derived from our radial velocity study 
 is  4.81 $\pm$ 0.01 M$_{\odot}$.   
 Using $q$ = 0.21 (donor rotating synchronously), we get $M_{2}$  $>$ 1.48 \msun and $M_{1}$  $>$ 7.05 \msun.
  On the other hand,  if $q$  = 0.54 we derive $M_{2}$  $>$ 6.16 \msun and $M_{1}$  $>$ 11.41 \msun. These masses
  turns to be too high for the temperatures derived from spectroscopy and this fact supports the $q$ = 0.21 solution.


To check if the formation of an accretion disc is possible, we calculate the distance to closest approach, measured
from the center of the gainer, of a stream coming from the inner Lagrangian point $L_{1}$:\\

$r_{\rm{min}} = 0.0488 q^{-0.464} a $\hfill(11)\\

\noindent
Lubow \& Shu (1975). For $q$ = 0.21 we get $r_{\rm{min}}$ = 0.10 $a$. i.e. 6.1 \rsun (0.065 $a$ or 4.0 \rsun for $q$ = 0.54).
When comparing with the gainer radius of 5.4 $R_{\odot}$ 
we observe that for the low mass ratio solution the disk can be formed ($r_{\rm{min}}$ $>$ $R_{1}$) but not for the
high mass ratio case, when an impact system should be observed where the gas stream directly impacts the gainer.

\section{Light curve model and system parameters}

\subsection{The fitting procedure}

The light-curve fitting was performed using the Nelder-Mead simplex algorithm (see e.g. Press et al. 1992) with optimizations described by Dennis and Torczon (1991), and the model of a binary system with a disc described in the previous section. 
For more detail see e.g. Djura\v{s}evi\'{c} (1992).

To obtain reliable estimates of the system parameters, a good practice is to restrict the number of free parameters by fixing some of them to values obtained from independent sources. 
In this Section we use subindexes $1$ and $2$ for labeling parameters of the hot and cool star, respectively.  
We fixed the mass ratio to $q$ = 0.21 and the stellar temperatures  to $T_{1}$ = 18.000 $K$ and $T_{2}$ = 8000 $K$ based on our spectroscopic results.  The hotter temperature was
selected to provide a good fit to the spectral energy distribution 
as explained in Section 6. The implications of this choice are discussed at the end of  Section 5.2.
In addition, we set the gravity darkening coefficient and the albedo of the gainer and the donor to $\beta_{1,2}$  = 0.25 and $A_{1,2}$ = 1.0 in accordance with von Zeipel's law for radiative envelopes (Von Zeipel 1924) and complete re-radiation (Rafert \& Twigg 1980). The limb-darkening for the components was calculated in the way described by Djura\v{s}evi\'{c} et al. (2010).

The possible values of free parameters are constrained by imposing the lowest and highest values which seem reasonable based on previous studies of this binary. Here are the ranges for the fitted parameters:

\begin{itemize}

\item Inclination: 50.0 to 70.0 degrees.
\item Disk dimension factor (the ratio of the disk radius and the radius of the critical Roche lobe along the y-axis): 0.5 to 0.9.
\item Disk edge temperature: 4000 to 8000 K.
\item Disk edge thickness: 0.02 to 0.06 (in units of $a_{\rm orb}$).
\item Disk center thickness: 0.13 to 0.17 (in units of $a_{\rm orb}$).
\item The exponent of the disk temperature distribution: 6.0 to 8.0.

\end{itemize}

After the first fit, these ranges were decreased according to the results of the first iteration.

We treated the rotation of the donor as synchronous ($f_{2}$ = 1.0), 
since it is assumed that the donor has filled its Roche lobe (i.e. the filling factor of the donor was set to $F_{2}$ = 1.0). Although it is expected that the accreted material from the disc would transfer enough angular momentum to increase the rotation rate of the gainer to the critical velocity (Packet 1981, de Mink, Pols \& Glebbeek 2007), our study cannot discriminate between synchronous and non-synchronous gainer, probably because it is partly hidden by the  accretion disc and rotationally  sensitive absorption lines are produced in the disc more that in the gainer. For this reason we present both solutions in this paper; they practically do not differ in physical parameters. 

We were able to model the asymmetry of the light curve very precisely by incorporating two regions of enhanced radiation on the disc: the hot spot (hs), and the bright spot (bs). The hot spot and the bright spot in our model are located on the edge side of the disk and are described by the longitude of the center of the spot, the angular  dimension of the spot, and the temperature ratio of the spot and  the unperturbed local temperature of the disk. The difference between the temperature of the spot and the local unperturbed temperature of the disk is what results in the difference in brightness. Location of the hot spot is calculated from the assumption that the gas stream from the L1 point falls tangentially onto the disc. The bright spot can be located at any longitude. The angular  dimension of the spots was constrained to the range from 10 to 40 degrees for the hot spot, and from 10 to 90 degrees for the bright spot; the temperature ratio for the spots can be from 1.0 to 2.0.  The incorporation in the model, of extended spots at the disk outer rim,  follows results of hydrodynamical simulations of gas dynamics in interacting close binary stars showing similar structures (e.g. Bisikalo et al. 2003), as explained in detail in the next section.

\subsection{The best light-curve model}

The  fit, $O-C$ residuals, individual flux contributions of the donor, disc and the gainer, and the view of the optimal model at orbital phases 0.25, 0.50 and 0.75, obtained with the parameters estimated by the light curve analysis, are illustrated in Fig.\,10 for the gainer's synchronous case. We note that residuals show no dependence on orbital or long-cycle phases
and that the best fit model of HD\,170582 contains an optically and geometrically thick  disc around the hotter, more massive gainer star. Our results for the synchronic gainer are shown in Table\,7 and those for a gainer rotating at critical velocity in Table\,8. Small differences are found in the physical parameters of both cases. 
It is reasonable to assume that the true parameters of the system are found in between both solutions.
For simplicity we discuss here the synchronous case only.

The best model shows that the inclination angle is well constrained to 67.4 $\pm$ 0.4 degree. With a radius of $R_{d} \approx 20.8 R_{\odot}$, the disc is 3.8 times larger than the central star ($R_{h} \approx 5.5 R_{\odot}$). The disc has a convex shape, with central thickness $d_{c} \approx 9.5 R_{\odot}$ and edge thickness $d_{e} \approx 2.3 R_{\odot}$. The temperature of the disc increases from $T_{d}$ = 5410 $K$ at its edge, to $T_{h}$ = 18000 $K$ at the inner radius, where it is in thermal and physical contact with the gainer. The relatively large disc temperature gradient explains the big difference between disc thickness at the inner and outer edges. In our model the gainer rotates synchronously with lineal velocity $v_{\rm{1r}} = v_{\rm{2r}} (R_{1}/R_{2}$) = 15.5 km s$^{-1}$. 
The surface gravity for the giant is larger than the figure obtained from the spectroscopic analysis (\logg = 2.3 $\pm$ 0.1 versus \logg = 1.50 $\pm$ 0.25).

In the best model the hot spot with 19.6\dg angular  dimension covers 10.9\% of the visible disc outer rim and it is situated at longitude $\lambda_{hs} \approx$  334\dg, roughly between the components of the system, at the place where the gas stream falls onto the disc (Lubow \& Shu 1975). 
The temperature of the hot spot is approximately 66\% higher than the disc edge
temperature, i.e. $T_{hs} \approx 9000$ $K$.

Although including the hot spot region into the model improves the fit, it cannot explain the light curve asymmetry completely. By introducing one additional bright spot, larger than the hot spot and located on the disc edge at $\lambda_{bs} \approx$ 135\dg, the fit becomes much better, i.e. has a lower $\chi^{2}$. This bright spot has $T_{bs} \approx$ 7900 $K$ and with 56.2\dg  angular dimension covers 31.2\% of the visible disc outer rim.

The hot and bright spots might be tentatively identified with shock regions,  characterized by higher density and higher temperature than the surrounding medium, revealed in hydrodynamical simulations of mass transfer in close binaries by Bisikalo et al. (1998, 1999, 2003). In particular, 
the hot spot is near the place where a ballistic trajectory of a particle released in the inner Lagrangian point intersects the accretion disc.  The bright spot  could correspond to the {\it hotline},
a shock region that, according to Bisikalo et al. (2003),  should appear as product of the interaction of the circum-disk halo and the stream.

 It is interesting to determine where the light centers of helium absorption components C1 and C2 calculated in Section 4.3 are located in the system.
The C1 light center is on the disc outer rim in the second quadrant 
and C2 is roughly in the direction of the hot spot 
but  far from the disc outer rim. Actually, the large $r$ = 45.9 \rsun indicates a position inside the donor for C2.
This is inconsistent with the donor temperature which is not enough for forming helium absorption. For the same reason, an
origin at the base of the gas stream is also hard to accept. However, we notice that the positions for the components were determined assuming Keplerian orbits (Eq.\,8). Therefore,
a possible interpretation for the puzzling result is the existence of  vertical motions in the hot spot region,  a wind where \textsc{He\,i\,5875}  absorption occurs, characterized 
by high temperature (T $\geq$ 10\,000) and projected velocities larger than Keplerian. This wind is really expected in the stream/disc interaction region according to models of interacting close binaries (Deschamps et al. 2013,  van Rensbergen et 
al. 2008).

 We find a gainer mass of 9 $M_{\odot}$ too high for a temperature of 18\,000 $K$. This B-type dwarf should have 
a temperature of 21\,000 $K$ (Lang 1999). It is possible that the gainer temperature found in the analysis of the spectral energy distribution (Section 6) and used in our model
is biased to low temperatures because of the presence of the low-temperature circumprimary disc and the low visibility of the gainer, which is almost completely hidden by the disc. 
To explore the sensitivity of our solution to the temperature of the gainer, we searched for the best synchronous solution with $T_{1}$ = 21\,000 $K$,
and we found basically the same parameters that for the cooler gainer case (Table\,9). Our conclusion is that the best solution is almost insensitive to the choice of the gainer temperature
among reasonable values for early B-type stars,
and that the parameters found in our analysis are robust in this sense. 

\begin{figure*}

\includegraphics[angle=0,width=17cm]{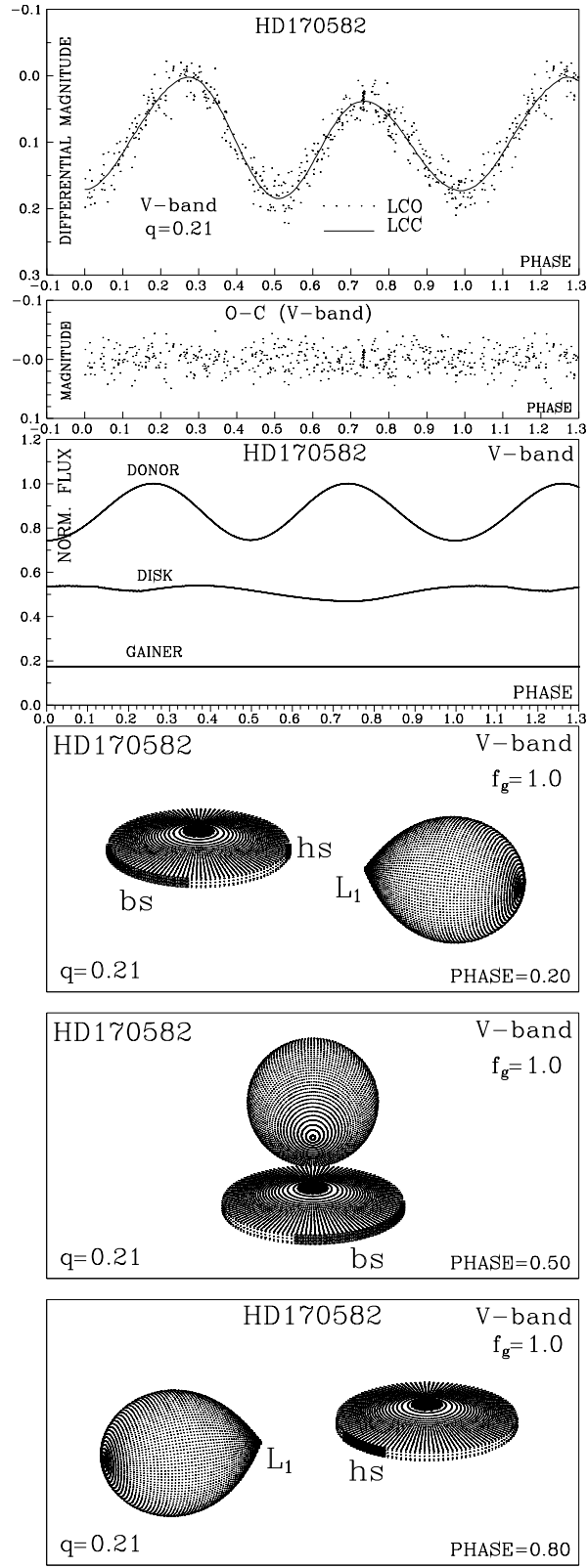}
\caption{Observed (LCO) and synthetic (LCC) light-curves of
{$\rm HD170582$} obtained by analyzing V-band photometric observations; final
O-C residuals between the observed and optimum synthetic light curves;
fluxes of donor, gainer and of the accretion disk, normalized
to the donor flux at phase 0.25; the views
of the optimal model at orbital phases 0.20, 0.50 and 0.80,
obtained with parameters estimated by the light curve analysis.}
\label{figV-s}
\end{figure*}

\section{Spectral energy distribution, reddening and distance}

We used the Spanish Virtual Observatory SED Analyzer\footnote{http://svo2.cab.inta-csic.es/theory/vosa4/}  (VOSA, Bayo et al. 2008) to get the  broad-band photometric fluxes published for HD\,170582 (Table\,10).
We performed a fit to the spectral energy distribution (SED) by means of the Marquant-Levenberg non-linear least square algorithm by  minimization of $\chi^{2}$ of the function:\\

\begin{small}
$f_{\lambda}=  f_{\lambda,0} 10^{-0.4E(B-V)[k(\lambda -V)+R(V)]}, $\hfill(12)\\
\end{small}

\noindent
where:\\

\begin{small}
$ f_{\lambda,0}= (R_{2}/d)^2 [(R_{1}/R_{2})^{2} f_{1, \lambda} +  f_{2, \lambda}], $\hfill(13)\\
\end{small}

\noindent 
and  $f_{1}$ and $f_{2}$ are the fluxes of the primary and secondary star,  $k(\lambda-V) \equiv E(\lambda-V)/E(B-V)$ is the normalized extinction
curve, $R(V) \equiv A(\lambda)/E(B-V)$ is the ratio of reddening to extinction at $V$, $d$ is the distance to the binary and $R_{1}/R_{2}$ is the ratio of the primary radius to the secondary radius. We used the average Galactic Extinction Curve parametrized by Fitzpatrick \& Massa (2007, hereafter FM07) to calculate reddened fluxes. The code was implemented in ORIGIN\footnote{http://www.originlab.com}. The stellar fluxes were taken from the grid 
of ATLAS9 Kurucz ODFNEW /NOVER models available in the Theoretical Spectra Web Server of the Spanish Virtual Observatory\footnote{http://svo2.cab.inta-csic.es/theory/newov/}. We used fluxes calculated with solar chemical abundance and microturbulence velocity 2 km s$^{-1}$. The free parameters of the fit were $R_{2}/d$ and $E(B-V)$. We fixed $R$ = 3.0 (FM07),  $\log g_{1}$= 4.0, $T_{2}$= 8000 K, $\log g_{2}$= 1.5
and $R_{1}/R_{2}$ = 0.346.  We tried models with temperatures $T_{1}$ = 15 kK, 18 kK, 20 kK and 22 kK. 
The deviating points from Lahulla and Hilton (1992) at $\lambda$ 3650 \AA, and that of the DENIS survey at $\lambda$ 7862 \AA\ were not considered in the fit. While the large deviation of the last one suggests an instrumental error, the first one could indicate a diminished Balmer jump regarding a normal star, as seen in some Be stars, a fact that is generally associated to the effect of a circumstellar envelope (Goraya 1986). It is important to keep in mind that the results of this section are limited to the validity of using the average Galactic extinction, which is not always 
true for different line of sights of our Galaxy (FM07). However, it is the only approximation possible at this moment.  Another limitation of the model is the absence of the
circumprimary disc component.

The best fit, minimizing $\chi^{2}$,  gave  $T_{1}$ = 18000 $\pm$ 1500 $K$, $R_{2}/d$= (1.478 $\pm$ 0.045) $\times$ 10$^{-9}$ and $E(B-V)$= 1.387 $\pm$ 0.015
(Fig.\,11).  As noted before $T_{1}$ is too low for the gainer mass derived in Section 5.2. This could indicate that we are fitting  the flux of the gainer plus the surrounding optically thick accretion disc, a pseudo-photosphere with a stellar-like flux distribution and temperature lower than the gainer. A disc surrounding a hot gainer and mimicking a cooler star is observed in the interacting binaries RX Cassiopeiae, W Crucis and W Serpentis (Plavec 1992); the possible connection of HD\,170582 with W Serpentis stars is discussed in Section 7. 

The relatively large extinction matches
published  values of Galactic dust reddening and extinction in the region of HD\,170582, viz.\, $E(B-V)$ = 1.37 $\pm$ 0.06 and 1.60 $\pm$ 0.07 (NASA/IPAC infrared science archive, based on Schlegel, Finkbeiner \& Davis 1998 and Schlafly and Finkbeiner 2011). 
Our color excess differs from that derived by Lahulla \& Hilton (1992), viz.\, $E(B-V)$ = 0.17, but they assumed a luminosity class V and a single A9 star in their analysis.  

Although the color excess is relatively well constrained from the SED analysis, the strength of diffuse interstellar bands (DIBs) 
indicate a different value. These bands are  absorption lines observed in the optical and infrared spectra of reddened stars (Herbig 1995); the strength
of some of them roughly correlated with the color excess produced by interstellar reddening.
We measured the equivalent widths ($EWs$) of DIBs located at 5780, 5797 and 8620 \AA\, ($EW$ = 0.30 $\pm$ 0.01, 0.08 $\pm$ 0.01 and 0.20 $\pm$ 0.01 \AA, respectively)
and used the relations given by  Munari (2000) and  Weselak et al. (2008) to estimate $E(B-V)$ = 0.55 $\pm$ 0.15. On the other hand,
the strength of \textsc{K\,i\,7699} suggests $E(B-V) \approx$ 0.5 (Munari \& Zwitter 1997). 

A possible explanation for this discrepancy is the existence of intrinsic reddening produced by circumstellar matter, such as $E(B-V) = E(B-V)_{\rm{is}} + E(B-V)_{\rm{cm}}$,
where $E(B-V)_{\rm{is}}$ = 0.55 is the interstellar reddening  and $E(B-V)_{\rm{cm}}$ = 0.82 is the circumstellar reddening. 
The existence of anomalous diffuse interstellar bands, weaker than expected for the stellar reddening,
has been reported for the Be stars HD\,44458 and HD\,63462 (Porceddu et al. 1992) and the Herbig Be star HD\,53367  (Whittet \& Blades 1980) and attributed to circumstellar matter.
The idea behind is that the carriers of diffuse bands (whatever they are), cannot survive in the relatively dense regions of circumstellar shells (Porceddu et al. 1992).  If this were the case for HD\,170582, then its relatively large reddening  should not be related to the location of the system in the  molecular cloud L\,379, but to the presence of circumstellar matter. Actually, the location of the system in a dense interstellar environment seems not to be related to their nature of Double Period Variable, since many of them exist in not so dense regions.
 
 From $R_{2}/d$= (1.478 $\pm$ 0.045) $\times$ 10$^{-9}$ determined from the SED fitting and using $R_{2}$ = 15.6 $\pm$ 0.2 $R_{\odot}$ determined from the LC model we obtain
 a distance of 238 $\pm$ 10 pc.  This figure compares well with the distance 210 pc derived by Lahulla \& Hilton (1992). Results of this section are summarized in Table\,11.
 


\begin{figure}
\scalebox{1}[1]{\includegraphics[angle=0,width=8.5cm]{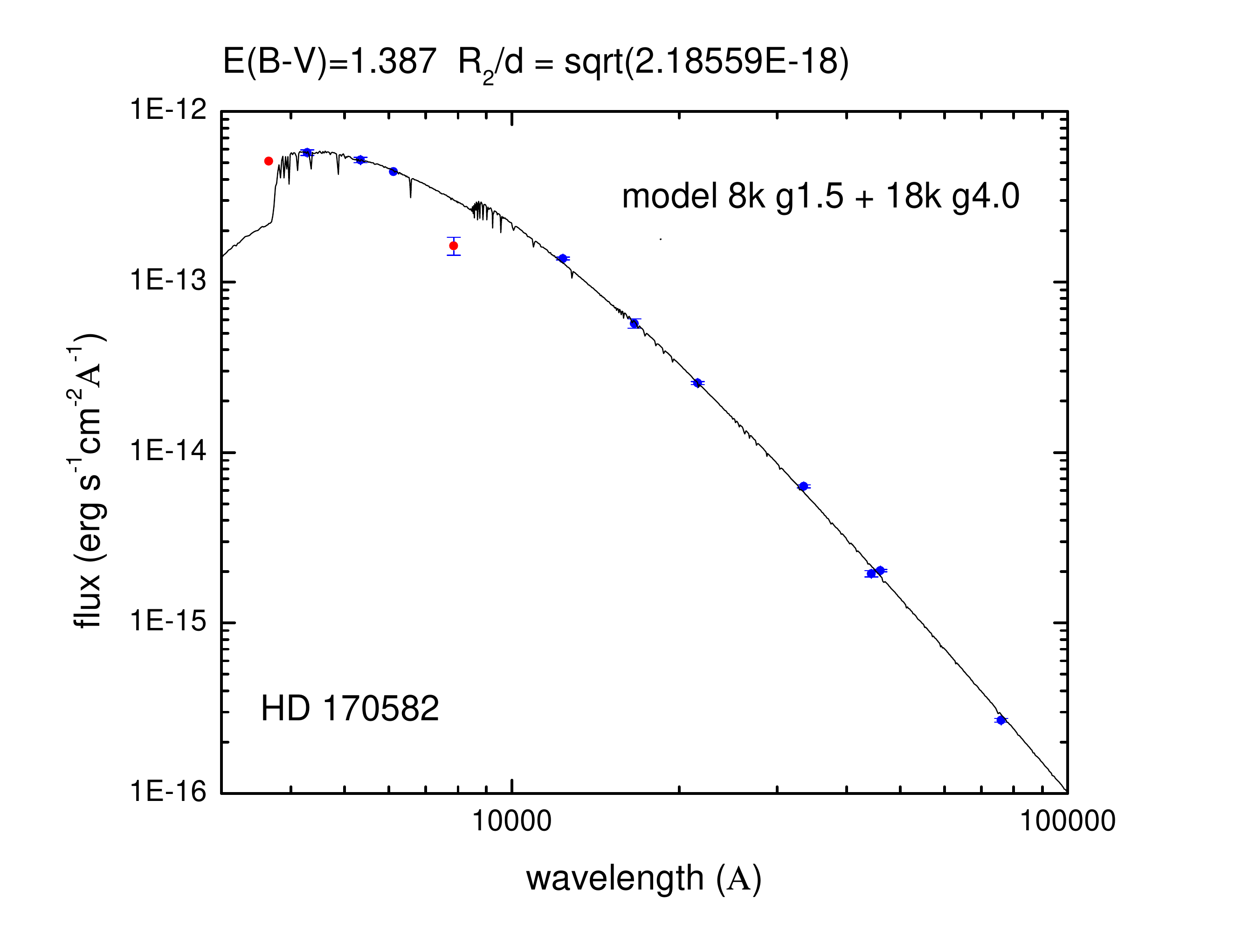}}
\caption{Spectral energy distribution and the best fit given by Eq.\,12, excluding the two outliers discussed in the text.}
  \label{x}
\end{figure}

\section{Conclusions}

We have investigated spectroscopically and photometrically the Double Period Variable HD\,170582.
It turns to be an interacting binary  consisting of a 8000 $K$ supergiant of solar abundance transferring matter 
to an early B-type star. From a radial velocity study based on high-resolution spectra, we find a mass function of $f(M)$ = 4.81 $\pm$ 0.01 \msun. Under the reasonable assumption of a donor with spin-orbit  synchronization filling its Roche-lobe, we derive a system mass ratio of $q$ = 0.21. We model the light curve including synthetic  stellar fluxes and an optically thick accretion disc around the B-type star. Using an inverse-problem solving algorithm, we derive 
the  system inclination, stellar masses, radii, temperatures, surface gravities and bolometric luminosities along with the properties of the disc (e.g. the radial and vertical extension and the temperature as a function of the radial coordinate).
All these parameters along with the donor rotational velocity, are given in Tables\,7, 8, 9 and 11. 

The disc is luminous, contributing between 30\% and 40\% to the system luminosity at the V-band, depending on the orbital phase. 
 The gainer is almost completely hidden by the disc and its contribution to the total light is only 10\%; their temperature results low for the stellar mass probably due to this handicap. However, we show that our system parameters are robust for a range of reasonable gainer temperatures. Our study indicates that for HD\,170582, and possibly for others DPVs with luminous discs, the optical and infrared flux is dominated by the donor and the disc, a fact that should be taken into account when fitting the  spectral energy distribution in this wavelength range.
 The best model of the accretion disk includes two relatively hot bright spots in the outer disc rim, in opposite positions, 
whose properties are given in Tables\,7 and 8. One of these spots is located at the region where the gas stream hits the accretion disc and the other could be explained as a shock region as indicated by 
previous simulations of gas dynamics in close binary systems. 

 We find that HD\,170882 shows spectroscopic properties similar to other DPVs like AU\,Monocerotis, V\,393 Scorpii, DQ\,Velorum  and OGLE\,05155332-6925581 (Mennickent et al, 2008, 2012a, 2012b, Barr{\'{\i}}a et al. 2013, 2014, Garrido et al. 2013). These properties include the existence of an optically thick disc surrounding a B-type star and evidence for stream-disc interaction.
In this sense, HD\,170582 could be related to the strongly interacting binaries  of the W Serpentis type (e.g. Plavec \& Koch 1978; Plavec 1980, 1992), however we observe in HD\,170582 the distinctive characteristics of DPVs which place them apart from the W Serpentis group, namely the relatively constant orbital period (it is variable in the Serpentids), and the presence of a long photometric periodicity lasting about 33 times the orbital period. To our knowledge, these last two features have never been reported simultaneously in any W Serpentis star.  For instance,
the eclipsing W Serpentis star RX\,Cas also shows a primary hidden by an accretion disc (Andersen et al. 1989, Djura\v{s}evi\'{c} 1992), and unequal maxima in the orbital light curve (Gaposchkin 1944), but it shows a variable orbital period of 32.32739 days and $dP_{o}/dt \sim 10^{-7}$ (Pustylnik et al. 2007) and a long photometric cycle of 516.06 days (Gaposchkin 1944). 

 Among the DPVs so far studied, HD\,170582 is unique in showing a double  \textsc{He\,i\,5875} line; two absorption lines move in antiphase during the orbital cycle.  The light center of one of these absorptions is located in the disc outer rim in the second quadrant, and the other roughly in the direction of the hot spot, but well far from the disc outer edge and inside the donor. This position is contradictory with formation of helium absorption lines, hence we suggest that significant vertical motions are present near the hot spot producing a radial velocity half-amplitude larger than expected for Keplerian motion. This wind is predicted by models of interacting close binaries and could be a mechanism of mass and angular momentum loss in these systems (Deschamps et al. 2013, van Rensbergen et 
al. 2008).

From the SED analysis we find a distance of 238 $\pm$ 10 pc and a relatively large color excess, compatible
with reported average measurements in the field.  However, DIBs suggests a lower color excess. This might be explained by the presence of 
circumstellar matter.

We will investigate in a forthcoming paper the long-term variability, the properties of the circumstellar matter and the evolutionary stage of this system.

\section{Acknowledgments}

 We thank the anonymous referee who provided useful insights on the first version of this paper. This investigation is based on observations conducted under 
CNTAC proposals CN2012A-17 and CN2013A-91.
This publication makes use of VOSA, developed under the Spanish Virtual Observatory project supported from the Spanish MICINN through grant AyA2008-02156.
 This research has made use of the SIMBAD database,
operated at CDS, Strasbourg, France.
 R.E.M. acknowledges support by VRID-Enlace 214.016.001-1.0, Fondecyt 1110347 and the BASAL Centro de Astrof{\'{i}}sica y Tecnolog{\'{i}}as Afines (CATA) PFB--06/2007. 
 EN acknowledges the support from the NCN grant 2011/01/B/ST9/05448 and
1007/S/IAs/14 funds. Some of the calculations have been carried out in Wroc{\l}aw
Centre for Networking and Supercomputing (http://www.wcss.pl), grant No.~214. This research was funded in part by the Ministry of Education, Science and Technological Development of Republic of Serbia through the project ÓStellar PhysicsÓ (No. 176004). M.C. thanks the support of FONDECYT project 1130173 and Instituto de F{\'{i}}sica y Astronom{\'{i}}a de Valpara'so.

\begin{table*}
\centering
 \caption{Summary of spectroscopic observations. N is number of spectra. The HJD at mid-exposure for the first spectrum of the series is given. $\Phi_{\rm{o}}$ and $\Phi_{\rm{l}}$ refer to the orbital and long-cycle phase, respectively, and are calculated according to Eq.\,7 and Eq.\,1.}
 \begin{tabular}{@{}cccccccc@{}}
 \hline
UT-date &Observatory/Telescope & Instrument &N & exptime (s)  &HJD &$\Phi_{\rm{o}}$& $\Phi_{\rm{l}}$\\
\hline

2008-05-24&ESO/EULER &CORALIE&3 &1800 &2454610.882519 &  0.995   &	0.327    \\
2008-05-25&ESO/EULER &CORALIE&2&1800  &2454611.903408 & 0.056&	0.329 \\
2008-06-13&LCO/Du-Pont &Echelle &2 &900    &2454631.794399 & 0.235	 &0.363 \\
2009-05-13&LCO/Du-Pont &Echelle &1 &750     &2454965.917228  & 0.038&	0.932 \\
2008-08-22&ESO/EULER &CORALIE&1&1200  &2454701.652578 & 0.375	 & 0.482 \\
2008-08-23&ESO/EULER &CORALIE&2&1200 &2454702.665820 & 0.435&	0.483 \\
2008-08-24&ESO/EULER &CORALIE&5&1500 &2454703.630014 & 0.492 &	0.485\\
2009-08-24&LCO/Du-Pont &Echelle &4 &900  &2454998.638369 & 0.977&	0.988 \\
2009-06-15&LCO/Du-Pont &Echelle &4 &900  &2455068.596925  & 0.124& 	0.107 \\
2012-03-29&CTIO/1.5m&CHIRON &3&1200     &2456016.817323 & 0.324&	0.722 \\
2012-04-05&CTIO/1.5m&CHIRON &3&1200    &2456023.877949& 0.742& 	0.734 \\
2012-04-10&CTIO/1.5m&CHIRON &3&1200    &2456028.833956& 0.036 &	0.743 \\
2012-04-30&CTIO/1.5m&CHIRON &3& 1200   &2456048.725777 &0.215	& 0.777   \\
2012-05-05&CTIO/1.5m&CHIRON &3&1200    &2456053.891370&  0.521	&0.785 \\
2012-05-14&CTIO/1.5m&CHIRON &3& 1200    &2456062.708397&0.044& 	0.800 \\
2012-05-20&CTIO/1.5m&CHIRON &3&1200    &2456068.848760&0.408&	0.811 \\
2012-05-31&CTIO/1.5m&CHIRON &3& 1200    &2456079.881457& 0.062&  	0.830 \\
2012-06-04&CTIO/1.5m&CHIRON&6& 1200    &2456083.772875&    0.292&	0.836 \\
2012-06-10&CTIO/1.5m&CHIRON &3&1200    &2456089.834323& 0.651	&0.847\\
2012-06-14&CTIO/1.5m&CHIRON &2&1200    &2456093.762810& 0.884&	0.853 \\
2012-06-26&CTIO/1.5m&CHIRON &3&1200    &2456105.720519 & 0.593&	0.874 \\
2012-08-11&CTIO/1.5m&CHIRON &3&1200    &2456151.703993 &  0.318	&0.952 \\
2012-08-25&CTIO/1.5m&CHIRON &1&1800    &2456165.550340&  0.139	&0.976 \\
2013-03-10&CTIO/1.5m&CHIRON &1&927      &2456362.918475&  0.837&	0.312 \\
2013-06-05&CTIO/1.5m&CHIRON &3&1200   &2456449.804977&0.987&	0.460\\
2013-06-13&CTIO/1.5m&CHIRON &2&1500    &2456457.834031&   0.462&	0.473 \\
2013-06-15&CTIO/1.5m&CHIRON &3&1200    &2456459.854115&  0.582  &0.477 \\
2013-06-17&CTIO/1.5m&CHIRON &3&1200    &2456461.827297&  0.699&	0.480 \\
2013-06-19&CTIO/1.5m&CHIRON &3&1200    &2456463.818915&  0.817 &	0.484 \\
2013-06-25&CTIO/1.5m&CHIRON &3&1200    &2456469.811141 & 0.172&	0.494 \\
2013-06-29&CTIO/1.5m&CHIRON &3&1200    &2456473.683746& 0.402 & 0.500\\
2013-06-30&CTIO/1.5m&CHIRON &1&241      &2456474.801975& 0.468&	0.502 \\
2013-07-03&CTIO/1.5m&CHIRON &3&1200    &2456477.678008& 0.639 &	0.507 \\
2013-07-07&CTIO/1.5m&CHIRON &3&1200    &2456481.694693 & 0.877&	0.514 \\
2013-07-11&CTIO/1.5m&CHIRON &3& 1200   & 2456485.709283&0.115& 	0.521 \\
2013-07-13&CTIO/1.5m&CHIRON &3&1200    &2456487.681643& 0.232&	0.524 \\
2013-07-15&CTIO/1.5m&CHIRON &3&1200   &2456489.676610& 0.350& 	0.528\\
2013-07-17&CTIO/1.5m&CHIRON &3&1200   &2456491.688161&  0.469 &	0.531 \\
2013-07-23&CTIO/1.5m&CHIRON &3&1200   &2456497.691100& 0.825& 	0.541 \\
2013-07-27&CTIO/1.5m&CHIRON &3&1200    &2456501.729047& 0.064&	0.548 \\
2013-08-01&CTIO/1.5m&CHIRON &3&1200   &2456506.680476&0.358& 	0.557 \\
2013-08-03&CTIO/1.5m&CHIRON &3&1200   &2456508.675606& 0.476&	0.560 \\
2013-08-05&CTIO/1.5m&CHIRON &3&1200   &2456510.636133 &0.592&  0.563\\
2013-08-10&CTIO/1.5m&CHIRON &3&1200   &2456515.651580 & 0.889&	0.572 \\
2013-08-12&CTIO/1.5m&CHIRON &3&1200    &2456517.687903&  0.010&	0.575 \\
2013-08-14&CTIO/1.5m&CHIRON &3&1200   &2456519.695122 & 0.129&	0.579 \\
2013-08-16&CTIO/1.5m&CHIRON &3&1200    &2456521.675552 &0.246&	0.582 \\
2013-08-20&CTIO/1.5m&CHIRON &3& 1200   &2456525.616642&  0.480&	0.589 \\

\hline
\end{tabular}
\end{table*}

\newpage

\begin{table*}
 \caption{ Average chemical abundances for the donor star. The standard deviation is given as the error when N, the number of parts/lines from which these abundances were derived,  is larger than 2. For all the other cases (N $\leq$ 2), the average error of all elements (0.23) is considered. The solar abundances are from Asplund et al. (2009).}
 \begin{tabular}{@{}crcr@{}}
 \hline
Element & N &$\log\epsilon_{(\rm El)}$   & $\log\epsilon_{(\rm El)}$ \\
(Z)         &       &HD\,170582        &Sun \\
\hline
   6   &   3   &  8.68   $\pm$ 0.41   &  8.43    \\
   8  &    4 &    8.25   $\pm$   0.15  &   8.69    \\
  11   &   5   &  7.05    $\pm$  0.09   &  6.24    \\
  12    &  7    & 7.46     $\pm$ 0.23    & 7.60    \\
  13   &   2   &  6.48     $\pm$ 0.23   &   6.45 \\
  14   &  16   &  7.82    $\pm$  0.23    & 7.51\\
  16    &  3    & 7.59     $\pm$ 0.37    & 7.12    \\
  20  &   17  &   6.34   $\pm$   0.23  &   6.34\\
  21  &   11  & 2.87    $\pm$  0.16    & 3.15\\
  22 &    46  & 4.87 $\pm$     0.18 &    4.95\\
  23  &   11  & 4.02   $\pm$   0.26   &  3.93   \\
  24  &   43 &  5.67   $\pm$   0.16   &  5.64\\
  25  &   13  & 5.60   $\pm$   0.20  &   5.43\\
  26  &  111  & 7.43    $\pm$  0.12    & 7.50\\
  27   &   3   &  5.70  $\pm$    0.08  &   4.99\\
  28 &    22 &    6.32 $\pm$    0.19 &    6.22\\
  29  &    2   &  4.22  $\pm$ 0.23     &  4.19\\
  30 &     2   &  4.22  $\pm$ 0.23    &   4.56\\
  39  &    7   &  2.14   $\pm$   0.19    & 2.21  \\
  40  &    7   &  2.58   $\pm$   0.22    & 2.58\\
  56  &    3    & 1.82   $\pm$   0.13    & 2.18\\
  60  &    2  &   1.48   $\pm$ 0.23     &  1.42\\
  63   &   1   & 0.92    $\pm$ 0.23       &0.52\\
 \hline  
\end{tabular}   
\end{table*} 

\newpage

\begin{table*}
 \caption{Radial velocities of the donor and their errors.}
 \begin{tabular}{@{}crrcrr@{}}
 \hline
HJD & RV (km s$^{-1}$) & error (km s$^{-1}$)&HJD & RV (km s$^{-1}$) & error (km s$^{-1}$)\\
\hline
2456016.817323  & 123.077  &  0.576   &             2456463.846797  &   -129.481 &      0.553      \\
2456016.831268  & 122.705 &   0.561  & 2456469.811141  &  124.138   &       0.466    \\ 
2456016.845214  & 122.151   & 0.563   &2456469.825082  &   124.197 &   0.484     \\
2456023.877949  & -138.305 &   0.918 &   2456469.839868 &   124.374  &  0.471      \\   
2456023.891893   &-138.747  &   1.076   &2456473.683746 &   78.948  &  0.483       \\
2456023.905839   &-139.238  &   0.881   &2456473.697687  &  78.744   &  0.551       \\
2456028.833956  & 31.288  &  0.777   & 2456473.711628    & 78.730    & 0.528   \\ 
2456028.847900  & 31.417   &     0.690   &2456474.801975   & 24.714    &0.667   \\
2456028.861845  & 32.741 &   0.777    &   2456477.678008  &  -105.880   &     0.477    \\   
2456048.725777  & 137.003   &  0.934     &2456477.691950    & -106.418 &      0.481     \\
2456048.739722  &  136.574  &   0.927   &2456477.705890   &  -106.300     &    0.454      \\
2456048.753667   &  135.962  &   1.421   & 2456481.694693  &  -101.678   &     0.421    \\ 
2456053.891370   & -19.690     &    0.872   &2456481.708634   &  -100.842  &   0.457     \\
2456053.905316   & -24.102 &   1.593    &   2456481.722574   & -99.809   &     0.394      \\  
2456053.919262    & -21.101    &   0.690     &2456485.709283   & 92.776    &    0.564       \\
2456062.708397    & 37.087 &   0.694   &   2456485.723224 &   93.303   &      0.526       \\
2456062.722342   & 39.364     &  0.791    & 2456485.737165  &   94.899 &   0.624   \\ 
2456062.736286    & 38.642   &   0.459   &  2456487.681643 &   139.918    &0.478   \\
2456068.848760  &  75.111     &    0.486    &   2456487.695584 &   140.020  &  0.501    \\   
2456068.862704  &  74.932    &    0.677      &2456487.709524  &   140.105   &  0.460     \\
2456068.876649 &   73.976   &     0.542   &2456489.676610  &   111.141    & 0.525      \\
2456079.881457   &  51.508  &      0.463   & 2456489.690549 &   110.660    &0.510    \\ 
2456079.895401  &  53.076  &      0.439   &2456489.704490  &   110.331 &   0.516     \\
2456079.909346  &  52.806  &      0.443   &    2456491.688161  &  24.047  &  0.436      \\   
2456083.772875    & 132.789    & 0.567     &2456491.702101  &  23.108      &   0.554       \\
2456083.786819    & 131.921     &0.498      &2456491.716043 &   23.022    &0.525        \\   
2456083.800763  &  132.396    & 0.561    & 2456497.691100  &   -127.555   &  0.510       \\
2456083.815193     &132.305    &0.597     &2456497.705041 &   -127.102    & 0.516      \\
2456083.829138   & 131.286   &  0.559      &   2456497.718983  &  -126.540    &   0.497     \\ 
2456083.843082  &  132.661  &  0.679       &2456501.729047   &  54.682 &   0.550        \\
2456089.834323 &   -111.254  &  0.838   &2456501.742986   &  55.338  &   0.561        \\   
2456089.834323  &   -111.254 &   0.842  &  2456501.756926  &  55.363    &    0.432       \\
2456089.834323  &  -111.254   &  0.847   &2456506.680476  &   106.396 &   0.603      \\
2456093.762810  &  -96.119   &  0.490   &    2456506.694416 &   106.082   &  0.546    \\ 
2456093.776754   &  -95.180 &   0.733    & 2456506.708355  &  105.110       &   0.566   \\
2456105.720519    & -77.838    &    0.442      &2456508.675606  &  17.626   &    0.641     \\   
2456105.734460  &  -78.346  &     0.478  &   2456508.689546   &  18.238  &    0.872       \\
2456105.748401   &  -79.092 &   0.464  &   2456508.703486   & 16.241   &     1.755     \\
2456151.703993   & 125.185       &  0.575      &   2456510.636133    &-76.643     &    0.494   \\ 
2456151.717933    &125.114         &  0.657       &2456510.650074    & -77.083   &     0.487      \\
2456151.731873   & 124.450         &   0.692         &2456510.664855    & -78.440  &      0.517       \\ 
2456165.550340  &   107.411    &   0.503   & 2456515.651580    &-93.099    &     0.426     \\
2456449.804977 &   -13.799   &   0.500   &2456515.665520     &-91.941   &     0.439       \\ 
2456449.818917  &  -12.957     &   0.276    &    2456515.686404    &-91.941&        0.440       \\
2456449.832859  &   -12.406    &   0.343     &2456517.687903    &5.802 &   0.468      \\
2456457.834031   &  29.909    &   0.484      &2456517.701843 &   5.918  &  0.465     \\ 
2456457.851445  &  28.075       &  0.643    &2456517.715784  &   7.180  &  0.627        \\
2456459.854115   &  -70.808    &   0.447    & 2456519.695122  &  101.865    & 0.443        \\   
2456459.868057 &   -71.100    &   0.392      &   2456519.709062 &   102.380    & 0.495       \\
2456459.881998  &  -72.264 &        0.640       &2456519.723001   &  102.046 &   0.490      \\
2456461.827297 &   -131.490    &          0.622      &  2456521.675552    & 139.573  &  0.481    \\     
2456461.841237  &   -133.005  &         0.577   &2456521.689492 &   139.634 &   0.510   \\
2456461.855179  &  -132.800       &     0.528   &2456521.703432   &  139.566  &  0.488     \\   
2456463.818915  &  -130.404   &     0.517    &   2456525.616642  &  15.008  &  0.572       \\
2456463.832856  &   -129.967  &      0.493     &2456525.630581  &  13.725    & 0.612     \\
2456525.644520  & 12.044  &   0.568  & & &  \\ 
\hline  
\end{tabular}   
\end{table*}  

\newpage



\begin{table*}
\centering
 \caption{Orbital elements for the donor of HD\,170582 obtained by minimization of
 the $\chi^{2}$ parameter given by Eq.\,(1) .
 The value $\tau^{*} = \tau- 2450000$ is given and also the maximum and minimum quantity in one isophote
 1$\sigma$. }
 \begin{tabular}{@{}lccc@{}}
 \hline
Parameter & best value & lower limit & upper limit \\
\hline
$P_{\rm{o}}$ (days)  &16.8722 & 16.8705 &16.8739 \\
$\tau^{*}$ &6029.56 &6027.90 &6030.96\\
$e$ &0.0133 &0.0055 &0.0205\\
$\omega$ &5.208 &5.183 &5.231\\
$K_{2}$ (km s$^{-1}$)&140.1 &139.0 &141.2 \\
$\gamma$ (km s$^{-1}$)&-1.30 &-2.05 &-0.55\\
\hline
\end{tabular}
\end{table*}

\newpage

\begin{table*}
 \caption{Radial velocities of helium lines with typical error 3 km s$^{-1}$.}
 \begin{tabular}{@{}cccrrr@{}}
 \hline
HJD & $\Phi_{\rm{o}}$ & $\Phi_{\rm{l}}$ &$RV_{5875}$ C1 (km s$^{-1}$ )&$RV_{5875}$ C2 (km s$^{-1}$ )  & $RV_{7065}$ (km s$^{-1}$) \\ 
\hline
2456016.817323 &0.324&0.722&-75.5& 100.9                &-  \\
2456016.831268 &0.325&0.722&-73.7& 121.7                &-  \\
2456016.845214 &0.325&0.722&-76.2&109.3               &-  \\
2456023.877949 &0.742&0.734&-&-121.7    &-       \\
2456023.891893 &0.743&0.734&-&	-125.8     &-  \\
2456028.833956 &0.036&0.743&49.0&-&-    \\
2456028.861845 &0.038&0.743&67.6&-&-  \\
2456151.703993	&0.318&	0.952&-&	134.0&-\\
2456151.717933	&0.319&	0.952&-&	133.0&-\\
2456151.731873	&0.320&	0.952&-&	135.6&- \\
2456165.550340  &0.139&	0.976&-&	106.4&-\\
2456449.804977 &0.987&0.460&38.8&-&40.7 \\
2456449.818917 &0.987&0.460&48.6&-&47.6  \\
2456449.832859 &0.988&0.460&46.3&-&49.6 \\
2456459.854115 &0.582&0.477&-&-58.8&-  \\
2456461.827297 &0.699&0.480&-&-106.3&-        \\
2456461.841237 &0.700&0.480&39.2&-&- \\
2456461.855179 &0.701&0.480&43.5&-&-  \\
2456463.818915 &0.817&0.484&116.8&-&-  \\
2456463.832856 &0.818&0.484&105.6&-103.4&- \\
2456463.846797 &0.819&0.484&102.6&-99.7&-  \\
2456469.811141 &0.172&0.494&-18.9&114.8&-    \\
2456469.825082 &0.173&0.494&-18.7&113.7&- \\
2456469.839868 &0.174&0.494&-12.3&135.7&-  \\
2456473.683746 &0.402&0.500&-51.9&100.5&-  \\
2456473.697687 &0.403&0.501&-54.8&100.0&- \\
2456473.711628 &0.404&0.501&-50.0&92.6&-  \\
2456477.678008 &0.639&0.507&-&-&27.6 \\
2456477.705890 &0.640&0.507&-&-&26.2  \\
2456481.694693 &0.877&0.514&-&-67.9&- \\
2456481.708634 &0.878&0.514&-&-74.7&-  \\
2456481.722574 &0.878&0.514&-&-67.2&-            \\
2456485.709283 &0.115&0.521&22.4&-&32.6 \\
2456485.723224 &0.115&0.521&30.0&-&-  \\
2456485.737165 &0.116&0.521&24.5&-&-  \\
2456487.681643 &0.232&0.524&-35.8&135.8&- \\
2456487.695584 &0.232&0.524&-30.5&141.1&-  \\
2456487.709524 &0.233&0.524&-36.0&138.4&-           \\
2456489.676610 &0.350&0.528&-52.3&116.8&-51.3 \\
2456489.690549 &0.351&0.528&-48.7&115.4&-57.1  \\
2456489.704490 &0.351&0.528&-47.3&118.6&-64.8  \\
2456497.691100 &0.825&0.541&113.3&-99.6&118.1 \\
2456497.705041 &0.826&0.541&117.4&-104.5&-  \\
2456497.718983 &0.826&0.541&114.9&-101.8&110.8  \\
2456501.729047 &0.064&0.548&31.0&-&- \\
2456501.742986 &0.065&0.548&27.5&-&-  \\
2456501.756926 &0.066&0.548&19.1&-&-               \\
2456506.680476 &0.358&0.557&-46.8&116.0&-54.4 \\
2456506.694416 &0.358&0.557&-45.0&113.9&-42.8  \\
2456506.708355 &0.359&0.557&-37.8&115.9&-  \\
2456515.651580 &0.889&0.572&92.8&-78.9&- \\
2456515.665520 &0.890&0.572&97.9&-76.8&-  \\
 2456515.686404 & 0.649&0.572&97.1&-75.0&-         \\
2456517.687903 &0.010&0.575&35.6&-&- \\
2456519.695122 &0.129&0.579&18.5&-&30.9 \\
2456519.709062 &0.130&0.579&8.8&-&30.0  \\
2456519.723001 &0.131&0.579&-&-&29.6         \\
2456521.675552 &0.246&0.582&-37.0&136.2&- \\
2456521.689492 &0.247&0.582&-31.1&142.3&-  \\
2456521.703432 &0.248&0.582&-&137.7&-  \\
2456525.616642 &0.480&0.589&-&137.7&- \\
\hline  
\end{tabular}   
\end{table*}  

\newpage

\begin{table*}
\centering
 \caption{Results of the sinusoidal fits ($\gamma + K \sin(2\pi(\Phi_{\rm{o}}-\delta)$)) to the RV curves of the \textsc{He\,i} 5875 components. The root mean square of the fits are also given. The parameters $\gamma$,
 $K$ and $rms$ are given in \kms.}
 \begin{tabular}{@{}lcccr@{}}
 \hline
Line &$\gamma$  &$K$  & $\delta$ &$rms$\\
\hline
\textsc{He\,i} 5875 C1 & 20.6 $\pm$ 2.3   &75.1 $\pm$ 2.7  & 0.601 $\pm$ 0.008  &15.3\\
\textsc{He\,i} 5875 C2 &   11.8 $\pm$ 1.4    &127.2 $\pm$  1.8&  0.004 $\pm$ 0.003 &7.9 \\
\hline
\end{tabular}
\end{table*}

\newpage


\begin{table*}

\caption{Results of the analysis of {$\rm HD170582$} V-band light-curve
obtained by solving the inverse problem for the Roche model with
an accretion disk around the more-massive (hotter) gainer in
synchronous rotation regime.}

 \label{TabbLyrae}
      \[
        \begin{array}{lrlr}
            \hline
            \noalign{\smallskip}

{\rm Quantity} & & {\rm Quantity} & \\
            \noalign{\smallskip}
            \hline
            \noalign{\smallskip}
   n                               & 455             & \cal M_{\rm_1} {[\cal M_{\odot}]} & 9.0  \pm 0.2  \\
{\rm \Sigma(O-C)^2}                & 0.1513          & \cal M_{\rm_2} {[\cal M_{\odot}]} & 1.9  \pm 0.1  \\
{\rm \sigma_{rms}}                 & 0.0182          & \cal R_{\rm_1} {\rm [R_{\odot}]}  & 5.5  \pm 0.2  \\
   i {\rm [^{\circ}]}              & 67.4  \pm 0.4   & \cal R_{\rm_2} {\rm [R_{\odot}]}  & 15.6 \pm 0.2  \\
{\rm F_d}                          & 0.65  \pm 0.02  & {\rm log} \ g_{\rm_1}             & 3.90 \pm 0.1  \\
{\rm T_d} [{\rm K}]                & 5410  \pm 200   & {\rm log} \ g_{\rm_2}             & 2.33 \pm 0.1  \\
{\rm d_e} [a_{\rm orb}]            & 0.155 \pm 0.004 & M^{\rm 1}_{\rm bol}               &-3.9  \pm 0.2  \\
{\rm d_2} [a_{\rm orb}]            & 0.038 \pm 0.004 & M^{\rm 2}_{\rm bol}               &-2.6  \pm 0.1  \\
{\rm a_T}                          & 7.3   \pm 0.3   & a_{\rm orb}  {\rm [R_{\odot}]}    & 61.2 \pm 0.2  \\
{\rm f_1}                          & 1.00            & \cal{R}_{\rm d} {\rm [R_{\odot}]} & 20.8 \pm 0.3  \\
{\rm F_1}                          & 0.187 \pm 0.004 & \rm{d_e}  {\rm [R_{\odot}]}       & 2.3  \pm 0.1  \\
{\rm T_1} [{\rm K}]                & 18000           & \rm{d_c}  {\rm [R_{\odot}]}       & 9.5  \pm 0.1  \\
{\rm T_2} [{\rm K}]                & 8000           &  & \\
{\rm A_{hs}=T_{hs}/T_d}            & 1.66  \pm 0.1   &                                                   \\
{\rm \theta_{hs}}{\rm [^{\circ}]}  & 19.6  \pm 2.0   &                                                   \\
{\rm \lambda_{hs}}{\rm [^{\circ}]} & 333.6 \pm 6.0   &                                                   \\
{\rm \theta_{rad}}{\rm [^{\circ}]} & 27.0  \pm 5.0   &                                                   \\
{\rm A_{bs}=T_{bs}/T_d}            & 1.46  \pm 0.1   &                                                   \\
{\rm \theta_{bs}} {\rm [^{\circ}]} & 56.2  \pm 3.0   &                                                   \\
{\rm \lambda_{bs}}{\rm [^{\circ}]} & 134.8 \pm 6.0   &                                                   \\
{\Omega_{\rm 1}}                   & 11.26 \pm 0.04  &                                                   \\
{\Omega_{\rm 2}}                   & 2.26  \pm 0.02  &                                                   \\
            \noalign{\smallskip}
            \hline
         \end{array}
      \]

FIXED PARAMETERS: $q={\cal M}_{\rm 2}/{\cal M}_{\rm
1}=0.21$ - mass ratio of the components, ${\rm T_1=18000 K}$ ;
${\rm T_2=8000 K}$  - temperature of the more massive (hotter)
gainer and less-massive (cooler) donor respectively, ${\rm F_2}=1.0$ -
filling factor for the critical Roche lobe of the donor,
$f{\rm _{1,2}}=1.00$ - non-synchronous rotation coefficients
of the system components, ${\rm \beta_{1,2}=0.25}$
- gravity-darkening coefficients of the components, ${\rm
A_{1,2}=1.0}$  - albedo coefficients of the components.

\smallskip \noindent Note: $n$ - number of observations, ${\rm
\Sigma (O-C)^2}$ - final sum of squares of residuals between
observed (LCO) and synthetic (LCC) light-curves, ${\rm
\sigma_{rms}}$ - root-mean-square of the residuals, $i$ - orbit
inclination (in arc degrees), ${\rm F_d=R_d/R_{yc}}$ - disk
dimension factor (the ratio of the disk radius to the critical Roche
lobe radius along y-axis), ${\rm T_d}$ - disk-edge temperature,
$\rm{d_e}$, $\rm{d_c}$,  - disk thicknesses (at the edge and at
the center of the disk, respectively) in the units of the distance
between the components, $a_{\rm T}$ - disk temperature
distribution coefficient, $f{\rm _g}$ - non-synchronous rotation coefficient
of the more massive gainer (in the synchronous rotation regime),
${\rm F_1}=R_1/R_{zc}$ - filling factor for the
critical Roche lobe of the hotter, more-massive gainer (ratio of the
stellar polar radius to the critical Roche lobe
radius along z-axis for a star in synchronous rotation regime),
${\rm A_{hs,bs}=T_{hs,bs}/T_d}$ - hot and bright spots' temperature
coefficients, ${\rm \theta_{hs,bs}}$ and ${\rm \lambda_{hs,bs}}$ -
spots' angular dimensions and longitudes (in arc degrees), ${\rm
\theta_{rad}}$ - angle between the line perpendicular to the local
disk edge surface and the direction of the hot-spot maximum
radiation, ${\Omega_{\rm 1,2}}$ - dimensionless surface potentials
of the hotter gainer and cooler donor, $\cal M_{\rm_{1,2}} {[\cal
M_{\odot}]}$, $\cal R_{\rm_{1,2}} {\rm [R_{\odot}]}$ - stellar
masses and mean radii of stars in solar units, ${\rm log} \
g_{\rm_{1,2}}$ - logarithm (base 10) of the system components
effective gravity, $M^{\rm {1,2}}_{\rm bol}$ - absolute stellar
bolometric magnitudes, $a_{\rm orb}$ ${\rm [R_{\odot}]}$,
$\cal{R}_{\rm d} {\rm [R_{\odot}]}$, $\rm{d_e} {\rm [R_{\odot}]}$,
$\rm{d_c} {\rm [R_{\odot}]}$ - orbital semi-major axis, disk
radius and disk thicknesses at its edge and center, respectively,
given in solar units.

\end{table*}

\begin{table*}

\caption{Results of the analysis of {$\rm HD170582$} V-band light-curve
obtained by solving the inverse problem for the Roche model with
an accretion disk around the more-massive (hotter) gainer in critical
non-synchronous rotation regime. Symbols are as in Table\,6. }

 \label{TabbLyrae}
      \[
        \begin{array}{lrlr}
            \hline
            \noalign{\smallskip}

{\rm Quantity} & & {\rm Quantity} & \\
            \noalign{\smallskip}
            \hline
            \noalign{\smallskip}
   n                               & 455             & \cal M_{\rm_1} {[\cal M_{\odot}]} & 9.0  \pm 0.2  \\
{\rm \Sigma(O-C)^2}                & 0.1542          & \cal M_{\rm_2} {[\cal M_{\odot}]} & 1.9  \pm 0.1  \\
{\rm \sigma_{rms}}                 & 0.0184          & \cal R_{\rm_1} {\rm [R_{\odot}]}  & 5.8  \pm 0.3  \\
   i {\rm [^{\circ}]}              & 67.4  \pm 0.4   & \cal R_{\rm_2} {\rm [R_{\odot}]}  & 15.6 \pm 0.2  \\
{\rm F_d}                          & 0.65  \pm 0.02  & {\rm log} \ g_{\rm_1}             & 3.86 \pm 0.1  \\
{\rm T_d} [{\rm K}]                & 5700  \pm 200   & {\rm log} \ g_{\rm_2}             & 2.33 \pm 0.1  \\
{\rm d_e} [a_{\rm orb}]            & 0.154 \pm 0.004 & M^{\rm 1}_{\rm bol}               &-4.0  \pm 0.2  \\
{\rm d_c} [a_{\rm orb}]            & 0.041 \pm 0.004 & M^{\rm 2}_{\rm bol}               &-2.6  \pm 0.1  \\
{\rm a_T}                          & 7.1   \pm 0.3   & a_{\rm orb}  {\rm. [R_{\odot}]}    & 61.2 \pm 0.2  \\
{\rm f_1}                          & 22.8  \pm 0.6   & \cal{R}_{\rm d} {\rm [R_{\odot}]} & 20.8 \pm 0.3  \\
{\rm F_1}                          & 1.00            & \rm{d_e}  {\rm [R_{\odot}]}       & 2.5  \pm 0.1  \\
{\rm T_1} [{\rm K}]                & 18000           & \rm{d_c}  {\rm [R_{\odot}]}       & 9.5  \pm 0.1  \\
{\rm T_2} [{\rm K}]                & 8000           &  & \\
{\rm A_{hs}=T_{hs}/T_d}            & 1.73  \pm 0.1   &                                                   \\
{\rm \theta_{hs}}{\rm [^{\circ}]}  & 19.0  \pm 2.0   &                                                   \\
{\rm \lambda_{hs}}{\rm [^{\circ}]} & 332.0 \pm 6.0   &                                                   \\
{\rm \theta_{rad}}{\rm [^{\circ}]} & 25.0  \pm 5.0   &                                                   \\
{\rm A_{bs}=T_{bs}/T_d}            & 1.43  \pm 0.1   &                                                   \\
{\rm \theta_{bs}} {\rm [^{\circ}]} & 56.0  \pm 3.0   &                                                   \\
{\rm \lambda_{bs}}{\rm [^{\circ}]} & 141.0 \pm 5.0   &                                                   \\
{\Omega_{\rm 1}}                   & 13.10 \pm 0.04  &                                                   \\
{\Omega_{\rm 2}}                   & 2.26  \pm 0.02  &                                                   \\
            \noalign{\smallskip}
            \hline
         \end{array}
      \]
\end{table*}

\begin{table*}

\caption{Results of the analysis of {$\rm HD170582$} V-band light-curve
obtained by solving the inverse problem for the Roche model with
an accretion disk around the more-massive (hotter) gainer in
synchronous rotation regime. Symbols are as in Table 6 but the gainer temperature is 21\,000 $K$. }

 \label{TabbLyrae}
      \[
        \begin{array}{llll}
            \hline
            \noalign{\smallskip}

{\rm Quantity} & & {\rm Quantity} & \\
            \noalign{\smallskip}
            \hline
            \noalign{\smallskip}
   n                               & 455             & \cal M_{\rm_1} {[\cal M_{\odot}]} & 9.0  \pm 0.2  \\
{\rm \Sigma(O-C)^2}                & 0.1516          & \cal M_{\rm_2} {[\cal M_{\odot}]} & 1.9  \pm 0.1  \\
{\rm \sigma_{rms}}                 & 0.0183          & \cal R_{\rm_1} {\rm [R_{\odot}]}  & 5.5  \pm 0.2  \\
   i {\rm [^{\circ}]}              & 67.4  \pm 0.4   & \cal R_{\rm_2} {\rm [R_{\odot}]}  & 15.6 \pm 0.2  \\
{\rm F_d}                          & 0.66  \pm 0.02  & {\rm log} \ g_{\rm_1}             & 3.9  \pm 0.1  \\
{\rm T_d} [{\rm K}]                & 5430  \pm 200   & {\rm log} \ g_{\rm_2}             & 2.33 \pm 0.1  \\
{\rm d_e} [a_{\rm orb}]            & 0.156 \pm 0.004 & M^{\rm 1}_{\rm bol}               &-4.5  \pm 0.2  \\
{\rm d_c} [a_{\rm orb}]            & 0.038 \pm 0.004 & M^{\rm 2}_{\rm bol}               &-2.6  \pm 0.1  \\
{\rm a_T}                          & 8.5   \pm 0.4   & a_{\rm orb}  {\rm [R_{\odot}]}    & 61.2 \pm 0.2  \\
{\rm f_1}                          & 1.00            & \cal{R}_{\rm d} {\rm [R_{\odot}]} & 21.2 \pm 0.3  \\
{\rm F_1}                          & 0.187 \pm 0.004 & \rm{d_e}  {\rm [R_{\odot}]}       & 2.4  \pm 0.1  \\
{\rm T_1} [{\rm K}]                & 21000           & \rm{d_c}  {\rm [R_{\odot}]}       & 9.6  \pm 0.1  \\
{\rm T_2} [{\rm K}]                & 8000           &  & \\
{\rm A_{hs}=T_{hs}/T_d}            & 1.85  \pm 0.1   &                                                   \\
{\rm \theta_{hs}}{\rm [^{\circ}]}  & 19.8  \pm 2.0   &                                                   \\
{\rm \lambda_{hs}}{\rm [^{\circ}]} & 330.6 \pm 6.0   &                                                   \\
{\rm \theta_{rad}}{\rm [^{\circ}]} & 10.0  \pm 8.0   &                                                   \\
{\rm A_{bs}=T_{bs}/T_d}            & 1.56  \pm 0.1   &                                                   \\
{\rm \theta_{bs}} {\rm [^{\circ}]} & 56.8  \pm 3.0   &                                                   \\
{\rm \lambda_{bs}}{\rm [^{\circ}]} & 103.7 \pm 6.0   &                                                   \\
{\Omega_{\rm 1}}                   & 11.25 \pm 0.04  &                                                   \\
{\Omega_{\rm 2}}                   & 2.26  \pm 0.02  &                                                   \\
            \noalign{\smallskip}
            \hline
         \end{array}
      \]

\end{table*}

\newpage

\begin{table*}
\centering
 \caption{Summary of broad-band photometric fluxes compiled from literature. Information about the DENIS public survey is found at http://cdsweb.u-strasbg.fr/denis.html.}
 \begin{tabular}{@{}ccccc@{}}
 \hline
Filter& $\lambda (\AA)$ &$f_{\lambda}$ (erg s$^{-1}$ cm$^{-2}$ $\AA^{-1}$)  &$ef_{\lambda}$ (erg s$^{-1}$ cm$^{-2}$ $\AA^{-1}$) &Reference\\
\hline
Johnson U  &3600.00 &5.120e-13 &-- &Lahulla and Hilton 1992\\
TYCHO/TYCHO.B&4280.00	 &  5.747e-13 &	2.064e-14&	Hog et al. (2000)\\		
TYCHO/TYCHO.V&5340.00  & 	5.198e-13 &	1.915e-14&	Hog et al. (2000)\\		
SLOAN/SDSS.r	      & 6122.33    &	4.428e-13 & --	&Adelman-McCarthy et al. (2008)\\		
DENIS/DENIS.I	& 7862.10   &	1.631e-13&   1.952e-14		&DENIS 3rd Release (Sep. 2005)\\	
2MASS/2MASS.J	& 12350.00 &	1.371e-13	 &  2.525e-15			&Skrutskie et al. (2006))\\
2MASS/2MASS.H	& 16620.00 &   5.710e-14 &	3.524e-15		&Skrutskie et al. (2006))\\	
2MASS/2MASS.Ks	& 21590.00 &	2.552e-14 &	5.642e-16		&Skrutskie et al. (2006))\\	
WISE/WISE.W1	& 33526.00  &	6.331e-15 &	1.283e-16		&Wright et al. (2010)\\
Spitzer/IRAC.I2  	& 44365.78  &   1.942e-15 &	7.778e-17		&Benjamin et al. (2003) and Churchwell et al. (2009)\\	
WISE/WISE.W2	& 46028.00	 &  2.025e-15 &	3.358e-17	&Wright et al. (2010)\\		
Spitzer/IRAC.I4	& 75891.59	  & 2.684e-16 &	6.548e-18	&Benjamin et al. (2003) and Churchwell et al. (2009)\\		
WISE/WISE.W3	& 115608.00  & 6.002e-17 &	1.161e-18	&Wright et al. (2010)\\		
WISE/WISE.W4	& 220883.00&  7.216e-18  &	9.638e-19  &Wright et al. (2010)\\
\hline
\end{tabular}
\end{table*}

\newpage

\begin{table*}
\centering
 \caption{Summary of light curve ephemerides, donor model spectrum and SED fit for HD\,170582. }
 \begin{tabular}{@{}cc@{}}
 \hline   
Parameter  &value  \\
\hline
Ephemeris$_{max, orbital}$  & 2\,452\,118.275 + 16.871 $\times\,E$   \\
Ephemeris$_{max, long}$  & 2\,452\,070.9 + 587 $\times\,E$ \\
$E(B-V)$                  &1.387 $\pm$ 0.015    \\
$d$                           &238 $\pm$ 10 pc  \\
$T_{1}$ &18000 $\pm$ 1500 K\\ 
$T_{2}$    &8000 $\pm$ 100 K \\
$v_{\rm{2r}}$ sin $i$ &44 $\pm$ 2 km s$^{-1}$\\
 \turb &1.0 $\pm$ 0.7 km s$^{-1}$. \\
log\,$g_{2}$ &1.7 $\pm$ 0.5 \\
\hline    
\end{tabular}
\end{table*}

\bsp 
\label{lastpage}
\end{document}